\documentclass[aps,prd,twocolumn,superscriptaddress,showpacs]{revtex4}
\usepackage{graphicx}% Include figure files
\usepackage{dcolumn}% Align table columns on decimal point
\usepackage{bm}% bold math
\usepackage{amssymb}  

\begin{document}

%Title of paper
\title{
A Study of \bm{$\bar{p}p\rightarrow$} Two Neutral Pseudoscalar Mesons
at the \bm{$\chi_{c0}(1^3P_0)$} Formation Energy}

% repeat the \author .. \affiliation  etc. as needed
% \email, \thanks, \homepage, \altaffiliation all apply to the current
% author. Explanatory text should go in the []'s, actual e-mail
% address or url should go in the {}'s for \email and \homepage.
% Please use the appropriate macro foreach each type of information

\affiliation{Fermi National Accelerator Laboratory, Batavia, Illinois 60510}
\affiliation{Istituto Nazionale di Fisica Nucleare and University of Ferrara, 44100 Ferrara, Italy}
\affiliation{Istituto Nazionale di Fisica Nucleare and University of Genoa, 16146 Genova, Italy}
\affiliation{University of California at Irvine, California 92697}
\affiliation{University of Minnesota, Minneapolis, Minnesota 55455}
\affiliation{Northwestern University, Evanston, Illinois, 60208}
\affiliation{Istituto Nazionale di Fisica Nucleare and University of Turin, 10125, Torino, Italy}
\author{M.~Andreotti}
\affiliation{Istituto Nazionale di Fisica Nucleare and University of Ferrara, 44100 Ferrara, Italy}

\author{S.~Bagnasco}
\affiliation{Istituto Nazionale di Fisica Nucleare and University of Genoa, 16146 Genova, Italy}
\affiliation{Istituto Nazionale di Fisica Nucleare and University of Turin, 10125, Torino, Italy}

\author{W.~Baldini}
\affiliation{Istituto Nazionale di Fisica Nucleare and University of Ferrara, 44100 Ferrara, Italy} 

\author{D.~Bettoni}
\affiliation{Istituto Nazionale di Fisica Nucleare and University of Ferrara, 44100 Ferrara, Italy} 

\author{G.~Borreani}
\affiliation{Istituto Nazionale di Fisica Nucleare and University of Turin, 10125, Torino, Italy} 

\author{A.~Buzzo}
\affiliation{Istituto Nazionale di Fisica Nucleare and University of Genoa, 16146 Genova, Italy}

\author{R.~Calabrese}
\affiliation{Istituto Nazionale di Fisica Nucleare and University of Ferrara, 44100 Ferrara, Italy}

\author{R.~Cester}
\affiliation{Istituto Nazionale di Fisica Nucleare and University of Turin, 10125, Torino, Italy}

\author{G.~Cibinetto}
\affiliation{Istituto Nazionale di Fisica Nucleare and University of Ferrara, 44100 Ferrara, Italy}

\author{P.~Dalpiaz}
\affiliation{Istituto Nazionale di Fisica Nucleare and University of Ferrara, 44100 Ferrara, Italy}

\author{G.~Garzoglio}
\affiliation{Fermi National Accelerator Laboratory, Batavia, Illinois 60510}

\author{K.~Gollwitzer}
\affiliation{Fermi National Accelerator Laboratory, Batavia, Illinois 60510}

\author{M.~Graham}
\affiliation{University of Minnesota, Minneapolis, Minnesota 55455}

\author{M.~Hu}
\affiliation{Fermi National Accelerator Laboratory, Batavia, Illinois 60510}

\author{D.~Joffe}
\affiliation{Northwestern University, Evanston, Illinois, 60208}

\author{J.~Kasper}
\affiliation{Northwestern University, Evanston, Illinois, 60208}

\author{G.~Lasio}
\affiliation{Istituto Nazionale di Fisica Nucleare and University of Turin, 10125, Torino, Italy}
\affiliation{University of California at Irvine, California 92697}

\author{M.~\surname{Lo~Vetere}}
\affiliation{Istituto Nazionale di Fisica Nucleare and University of Genoa, 16146 Genova, Italy}

\author{E.~Luppi}
\affiliation{Istituto Nazionale di Fisica Nucleare and University of Ferrara, 44100 Ferrara, Italy}

\author{M.~Macr\`\i}
\affiliation{Istituto Nazionale di Fisica Nucleare and University of Genoa, 16146 Genova, Italy}

\author{M.~Mandelkern}
\affiliation{University of California at Irvine, California 92697}

\author{F.~Marchetto}
\affiliation{Istituto Nazionale di Fisica Nucleare and University of Turin, 10125, Torino, Italy}

\author{M.~Marinelli}
\affiliation{Istituto Nazionale di Fisica Nucleare and University of Genoa, 16146 Genova, Italy}

\author{E.~Menichetti}
\affiliation{Istituto Nazionale di Fisica Nucleare and University of Turin, 10125, Torino, Italy}

\author{Z.~Metreveli}
\affiliation{Northwestern University, Evanston, Illinois, 60208}

\author{R.~Mussa}
\affiliation{Istituto Nazionale di Fisica Nucleare and University of Ferrara, 44100 Ferrara, Italy}
\affiliation{Istituto Nazionale di Fisica Nucleare and University of Turin, 10125, Torino, Italy}

\author{M.~Negrini}
\affiliation{Istituto Nazionale di Fisica Nucleare and University of Ferrara, 44100 Ferrara, Italy}

\author{M.~M.~\surname{Obertino}}
\affiliation{Istituto Nazionale di Fisica Nucleare and University of Turin, 10125, Torino, Italy}
\affiliation{University of Minnesota, Minneapolis, Minnesota 55455}

\author{M.~Pallavicini}
\affiliation{Istituto Nazionale di Fisica Nucleare and University of Genoa, 16146 Genova, Italy}

\author{N.~Pastrone}
\affiliation{Istituto Nazionale di Fisica Nucleare and University of Turin, 10125, Torino, Italy}

\author{C.~Patrignani}
\affiliation{Istituto Nazionale di Fisica Nucleare and University of Genoa, 16146 Genova, Italy}

\author{T.~K.~\surname{Pedlar}}
\affiliation{Northwestern University, Evanston, Illinois, 60208}

\author{S.~Pordes}
\affiliation{Fermi National Accelerator Laboratory, Batavia, Illinois 60510}

\author{E.~Robutti}
\affiliation{Istituto Nazionale di Fisica Nucleare and University of Genoa, 16146 Genova, Italy}

\author{W.~Roethel}
\affiliation{Northwestern University, Evanston, Illinois, 60208}
\affiliation{University of California at Irvine, California 92697}

\author{J.~L.~\surname{Rosen}}
\affiliation{Northwestern University, Evanston, Illinois, 60208}

\author{P.~Rumerio}
\affiliation{Istituto Nazionale di Fisica Nucleare and University of Turin, 10125, Torino, Italy}
\affiliation{Northwestern University, Evanston, Illinois, 60208}

\author{R.~Rusack}
\affiliation{University of Minnesota, Minneapolis, Minnesota 55455}

\author{A.~Santroni}
\affiliation{Istituto Nazionale di Fisica Nucleare and University of Genoa, 16146 Genova, Italy}

\author{J.~Schultz}
\affiliation{University of California at Irvine, California 92697}

\author{S.~H.~\surname{Seo}}
\affiliation{University of Minnesota, Minneapolis, Minnesota 55455}

\author{K.~K.~\surname{Seth}}
\affiliation{Northwestern University, Evanston, Illinois, 60208}

\author{G.~Stancari}
\affiliation{Fermi National Accelerator Laboratory, Batavia, Illinois 60510}
\affiliation{Istituto Nazionale di Fisica Nucleare and University of Ferrara, 44100 Ferrara, Italy}

\author{M.~Stancari}
\affiliation{University of California at Irvine, California 92697}
\affiliation{Istituto Nazionale di Fisica Nucleare and University of Ferrara, 44100 Ferrara, Italy}

\author{A.~Tomaradze}
\affiliation{Northwestern University, Evanston, Illinois, 60208}

\author{I.~Uman}
\affiliation{Northwestern University, Evanston, Illinois, 60208}

\author{T.~\surname{Vidnovic~III}}
\affiliation{University of Minnesota, Minneapolis, Minnesota 55455} 

\author{S.~Werkema}
\affiliation{Fermi National Accelerator Laboratory, Batavia, Illinois 60510}

\author{P.~Zweber}
\affiliation{Northwestern University, Evanston, Illinois, 60208}

\collaboration{Fermilab E835 Collaboration}
\noaffiliation


\begin{abstract}
Fermilab experiment E835 has studied reactions
$\bar{p}p \rightarrow \pi^0\pi^0,\,\pi^0\eta,\,\eta\eta,\,$ $\pi^0\eta'$ 
and $\eta\eta'$ in the energy region of the $\chi_{c0}(1^3P_0)$ from $3340$~MeV 
to $3470$~MeV.
Interference between resonant and continuum production 
is observed in the $\pi^0\pi^0$ and $\eta\eta$  
channels, and the product of the input and output branching 
fractions is measured.
Limits on resonant production are set for the $\pi^0\eta$ and 
$\pi^0\eta'$ channels.
An indication of interference is observed in the $\eta\eta'$ channel.
The technique for extracting resonance parameters
in an environment dominated by continuum production is described.
\end{abstract}

\maketitle

\section{\label{sec:intro}Introduction}

E835 has measured the cross section of antiproton-proton annihilation
into two pseudoscalar mesons ($\bar{p}p \rightarrow P_1P_2$)
in the energy region of the $\chi_{c0}$.
The final states studied are:
\begin{displaymath}
P_1P_2~\equiv~
\pi^0\pi^0~\mathrm{(1)},~
\pi^0\eta ~\mathrm{(2)},~ 
\eta\eta  ~\mathrm{(3)},~
\pi^0\eta'~\mathrm{(4)},~\mathrm{and}~
\eta\eta' ~\mathrm{(5)}.
\end{displaymath}
\setcounter{equation}{5} 
A total integrated luminosity of 33~pb$^{-1}$ has been collected at 
17 energy points from 3340~MeV to 3470~MeV.
The mesons were detected through their decay into
two photons. The cross sections reported here have been corrected for
the respective branching ratios~\cite{PDG}:
$B(\pi^0\rightarrow\gamma\gamma)=(98.798\pm0.032)\%$, 
$B(\eta\rightarrow\gamma\gamma)=(39.43\pm0.26)\%$, and 
$B(\eta'\rightarrow\gamma\gamma)=(2.12\pm0.14)\%$.

The analysis of reaction~(1) has been published in
letter format~\cite{2pi0}. 
In the present work, processes (2)-(5) 
are reported for the  first time 
and a more extensive discussion is given of process~(1).
More details on the analyses (1)-(3)
can be found in a dissertation~\cite{Paolo}.

The expression for the angular distribution of $\bar{p}p\rightarrow P_1P_2$, 
in the vicinity of the $\chi_{c0}$ resonance, is given in
Equations~(\ref{eq:withampres_AandB}),~(\ref{eq:interferingamp})~and~(\ref{eq:noninterferingamp}).
The subsequent isotropic decay of each meson into two photons 
need only to be considered for acceptance determination. 
\begin{equation}
\frac{d\sigma}{dz}
=\Big|-\frac{A_R}{x+i} + Ae^{i\delta_A} \Big|^2
+\Big| Be^{i\delta_B} \Big|^2,
\label{eq:withampres_AandB}
\end{equation}
\begin{equation}
A~e^{i\delta_A} \equiv
\sum_{J=0,2,4,\,...}^{J_{max}} 
(2J+1)~C_{J}~e^{i\delta_J}~P_{J}(z), \mbox{~and}
\label{eq:interferingamp}
\end{equation}
\begin{equation}
B~e^{i\delta_B} \equiv
\sum_{J=2,4,\,...}^{J_{max}} 
\frac{2J+1}{\sqrt{J(J+1)}}~C_{J}^{1}~e^{i\delta_J^1}~P_{J}^{1}(z).
\label{eq:noninterferingamp}
\end{equation}
The variable $z$ is defined as
\begin{equation}
z\equiv\vert\cos\theta^*\vert~, 
\label{eq:z}
\end{equation}
where $\theta^*$ is the production polar angle of the two mesons
in the center of mass (cm), while $x$ is defined as
\begin{equation}
x \equiv \frac{E_{cm}-M_{\chi_{c0}}}{\Gamma_{\chi_{c0}} /2}.
\label{eq:x}
\end{equation}
The term $Ae^{i\delta_A}$ is generated from the $\bar{p}p$~state with helicity 
$\lambda_i\equiv\lambda_{p}-\lambda_{\bar{p}}=0$ 
(helicity-0), where $\lambda_{p}$ and 
$\lambda_{\bar{p}}$ are the proton and antiproton helicities, respectively.
The helicity-0 coefficients and phases of the expansion are
$C_{J}$ and $\delta_{J}$, respectively, while  
$P_{J}(z)$ are Legendre polynomials.
%
%%%%%%%%%%%%%%%%%%%%%%%%%%%%%%%%%%%%%%%%%%%%%%%%%%%%%%%%%%%%%%%%%%%
\begin{figure}
\begin{center}
\includegraphics[width=21pc]{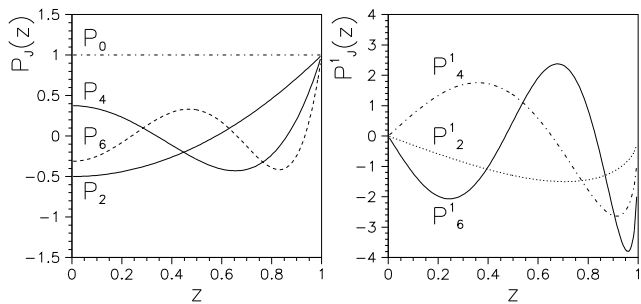}
\caption{ \label{fig:legendre_p0246}
The first Legendre polynomials $P_J^{M=0}\equiv P_J$ (left) 
and associated functions $P_J^{M=1}$ (right).}
\end{center}
\end{figure}
%%%%%%%%%%%%%%%%%%%%%%%%%%%%%%%%%%%%%%%%%%%%%%%%%%%%%%%%%%%%%%%%%%%
%
The term $ Be^{i\delta_B}$ is generated from the $\bar{p}p$~state with helicity 
$\vert\lambda_i\vert=1$ (helicity-1).
The helicity-1 coefficients and phases are 
$C_{J}^{1}$ and $\delta_{J}^{1}$, respectively, while
$P_{J}^{1}(z)$ are Legendre associate functions $P_{J}^{M}$ with $M=1$.
The $\chi_{c0}$ ($1^3P_0$) resonance, parameterized by a Breit-Wigner amplitude
$-A_R/(x+i)$, has been extracted from the $J=0$ term.
Both nonresonant terms, $Ae^{i\delta_A}$ and $ Be^{i\delta_B}$,
have a slowly varying (on the scale of the $\chi_{c0}$ width, $\approx 10$~MeV) 
dependence on $x$ and an angular dependence on $z$ described by the Legendre polynomials.
The amplitudes $-A_R/(x+i)$ and $Ae^{i\delta_A}$,
from the same helicity-0 initial state add coherently
and interfere with each other.
The amplitude $Be^{i\delta_B}$ is added incoherently because
it is generated from the helicity-1 initial state and does not interfere with 
the helicity-0 terms. 

Equation~(\ref{eq:withampres_AandB}) can also be written as
\begin{equation} 
\frac{d\sigma}{dz}=
\frac{A_R^2}{x^2+1}+A^2+
\underbrace{2 A_R A\,\frac{\sin{\delta_{A}}-x\cos{\delta_{A}}}{{x^2+1}}
}_{\mathrm{interference-term}}+B^2.
\label{eq:interfterm}  
\end{equation}
Given the moderate energy dependence of $Ae^{i\delta_A}$ and 
$Be^{i\delta_B}$, Equation~(\ref{eq:interfterm}) shows that at a fixed value 
of $z$, as $E_{cm}$ varies across the resonance, 
even a small resonant contribution can generate a significant interference
signal superimposed on the continuum. 
As an example, for a resonant amplitude $A_R$ that is one tenth 
of the helicity-0 continuum amplitude $A$, 
the peak contribution to the cross section from the Breit-Wigner term 
$A_R^2/(x^2+1)$ is only 1\% of $A^2$, 
while the factor $2A_R A$ of the interference
term of Equation~(\ref{eq:interfterm}) is 20\% of $A^2$.
The factor $(\sin{\delta_A}-x\cos{\delta_A})/(x^2+1)$ determines the 
shape of the interference pattern.
A small resonance and a large continuum is a typical experimental
condition found when studying charmonium in the process 
$\bar{p}p\rightarrow\mathrm{hadrons}$. 

The term $B^2$ provides an additional contribution to the
cross section, and if $B^2$ is too large compared to the interference
term, it may mask the presence of the resonance.
It is quite helpful that the associated Legendre functions
with $M=1$, which are the constituents of the amplitude $Be^{i\delta_B}$,
share a common multiplicative factor $z$ which 
causes them to vanish at $z=0$
(see Figure~\ref{fig:legendre_p0246}).
Since the polynomials $P_J^1$ are either squared or multiplied by 
each other, $B^2$ is suppressed with respect to $A^2$ by $z^2$ at small $z$.
To measure the value of the resonant amplitude $A_R$, it is critical to know 
the size of both the helicity-0 continuum, with which the resonance is coherent, 
and the helicity-1 continuum, which is incoherent with the resonance.
The region at small $z$ (i.e. $\cos\theta^*\approx 0)$ is therefore 
the natural place to concentrate for the determination of $A_R$.

\begin{table}
\caption{\label{tab:quantumnumbers}
How the orbital angular momentum $L$ and the spin $S$
of the $\bar{p}p$ in the initial state combine to make different $J^{PC}$ 
values and the corresponding charmonium resonances.
A $\checkmark$ indicates whether a pseudoscalar-pseudoscalar ($P_1P_2$) 
state is accessible. 
Spectroscopic notation is given for $\bar{p}p$ ($^{2S+1}L_{J}$) 
and $c\bar{c}$ ($n^{2S+1}L_{J}$).
}
\begin{center}
\begin{tabular}{|cc|c|cc|c|}\hline
 \multicolumn{2}{|c|}{$\bar{p}p$} &$      $& \multicolumn{2}{c|}{$\bar{c}c$} 
 &$P_1P_2                      $\\
$  L~~S       $&$^{2S+1}L_{J}   $&$J^{PC}$&resonance         
 &$n^{2S+1}L_{J}           $&$                            $\\
\hline
$  0~~0       $&$^{1}S_{0}      $&$0^{-+}$&$\eta_c~(\eta'_c)$
 &$1^{1}S_{0}~(2^{1}S_{0}) $&$                            $\\
$  0~~1       $&$^{3}S_{1}      $&$1^{--}$&$J/\psi~(\psi')  $
 &$1^{3}S_{1}~(2^{3}S_{1}) $&$                            $\\
\hline
$  1~~0       $&$^{1}P_{1}      $&$1^{+-}$&$h_c             $
 &$1^{1}P_{1}              $&$                            $\\
$  1~~1       $&$^{3}P_{0}      $&$0^{++}$&$\chi_{c0}       $
 &$1^{3}P_{0}              $&$\checkmark                  $\\
$  1~~1       $&$^{3}P_{1}      $&$1^{++}$&$\chi_{c1}       $
 &$1^{3}P_{1}              $&$                            $\\
$  1~~1       $&$^{3}P_{2}      $&$2^{++}$&$\chi_{c2}       $
 &$1^{3}P_{2}              $&$\checkmark                  $\\
\hline
$  2~~0       $&$^{1}D_{2}      $&$2^{-+}$&$                $
 &$                        $&$                            $\\
$  2~~1       $&$^{3}D_{1}      $&$1^{--}$&$\psi''          $
 &$1^{3}D_{1}              $&$                            $\\
$  2~~1       $&$^{3}D_{2}      $&$2^{--}$&$                $
 &$                        $&$                            $\\
$  2~~1       $&$^{3}D_{3}      $&$3^{--}$&$                $
 &$                        $&$                            $\\
\hline
$  3~~0       $&$^{1}F_{3}      $&$3^{+-}$&$                $
 &$                        $&$                            $\\
$  3~~1       $&$^{3}F_{2}      $&$2^{++}$&$                $
 &$                        $&$\checkmark                  $\\
$  3~~1       $&$^{3}F_{3}      $&$3^{++}$&$                $
 &$                        $&$                            $\\
$  3~~1       $&$^{3}F_{4}      $&$4^{++}$&$                $
 &$                        $&$\checkmark                  $\\
\hline
$  4~~0       $&$^{1}G_{4}      $&$4^{-+}$&$                $
 &$                        $&$                            $\\
$  4~~1       $&$^{3}G_{3}      $&$3^{--}$&$                $
 &$                        $&$                            $\\
$  4~~1       $&$^{3}G_{4}      $&$4^{--}$&$                $
 &$                        $&$                            $\\
$  4~~1       $&$^{3}G_{5}      $&$5^{--}$&$                $
 &$                        $&$                            $\\
\hline
\end{tabular}
\end{center}
\end{table}

Table~\ref{tab:quantumnumbers} shows how the orbital angular momentum 
($L_{\bar{p}p}$) and spin ($S_{\bar{p}p}$) combine to form different values of
the total angular momentum, parity and charge conjugation ($J^{PC}$).
For each $J^{PC}$, it shows which charmonium resonance is
formed, and whether the $J^{PC}=\mathrm{even}^{++}$ pseudoscalar-pseudoscalar 
final states can be accessed.
The spectroscopic notation ($^{2S+1}L_J$) is given for $\bar{p}p$ and 
$c\bar{c}$ states. Both states are fermion-antifermion, thus are described in 
the same way as far as $L$, $S$ and $J^{PC}$ are concerned, while a radial quantum number
($n$) applies only to the bound $c\bar{c}$ system.
The $\chi_{c0}$ ($J^{PC}=0^{++}$) can be produced only from the spin-triplet
$L_{\bar{p}p}=1$.
Table~\ref{tab:quantumnumbers} is sorted by increasing $L_{\bar{p}p}$, 
which correlates with the $\bar{p}p$ impact parameter ($b$).
Even for nonresonant reconfiguration of the $\bar{p}p$ system into two
pseudoscalar mesons, small impact parameter is favored at small $z$ 
since the valence quarks must either annihilate or suffer large momentum transfers.
Excellent fits to the angular distributions of the two pseudoscalar 
mesons are obtained with a limited number of partial waves.

\section{\label{sec:selection}Data Selection}

A comprehensive description of the E835 apparatus and experimental technique 
is provided in \cite{bigpaper}.
The data selection and the determination of the acceptance and efficiency 
for the processes (1)-(3) are extensively discussed in \cite{Paolo};
only a summary is given here.
Processes~(4) and~(5) are discussed in 
Section~\ref{sec:etaprime_cross_sections}.

The stochastically-cooled $\bar{p}$ beam circulating in the antiproton
accumulator intercepts a hydrogen gas-jet target. The energy of the
beam can be tuned to the energy of interest.
Before the year 2000 run, the accumulator transition energy 
was raised to $E_{cm}$ of 3600~MeV.
A technique was developed to modify the
accumulator lattice and lower the transition energy as the beam was
decelerated~\cite{McGinnis:gt}, thus allowing adequate margin 
between the operating energy
and the transition energy.
The $E_{cm}$ spectrum is approximately gaussian and is
determined from measurements of the 
$\bar{p}$-beam revolution frequency and orbit length.
The precision of the measurement of the mean value of the $E_{cm}$ spectrum
is about 100~keV. 
The r.m.s. spread of the $E_{cm}$  spectrum is a few hundred keV, 
much smaller than the width of the $\chi_{c0}$ resonance. 

The important detectors for this analysis are the central
calorimeter (CCAL) which is used to measure the photon energy deposits, 
the system of scintillation counters which vetoes events with charged particles
and the luminosity monitor.
The energy resolution of CCAL 
is $\sigma_E/E\simeq6\%/\sqrt{E(\mathrm{GeV})}+1.4\%$, 
while the polar and azimuthal angular resolutions 
are $\sigma_{\theta}\simeq 6$~mrad 
and $\sigma_{\phi}\simeq 11$~mrad, respectively. 

Online, two-body candidate events were selected by means of two
independent triggers based on CCAL: 
the two-body trigger and the total-energy trigger. 
The two-body trigger accepted events with two large 
energy deposits in CCAL satisfying two-body kinematics;
the  total-energy trigger
accepted events where at least 80\% of the center-of-mass energy was deposited in CCAL.
For $\pi^0\pi^0$ events, 
the efficiency of the two-body trigger is $99.9\%$~- the small 
opening angles of the decay 
$\pi^0\rightarrow\gamma\gamma$ keeps the 
$\pi^0\pi^0\rightarrow 2\gamma +2\gamma$ events within the two-body trigger
requirements - while the efficiency of the total-energy trigger is $\sim 98.2\%$. 
The two-body trigger is somewhat less efficient for 
$\eta\rightarrow\gamma\gamma$ decays and 
only the total-energy trigger was used for selecting 
$\pi^0\eta$ and $\eta\eta$
events, with an average efficiency of $\sim 98.5\%$ for both channels.
Both triggers were subjected to the charged-particle veto.

%%%%%%%%%%%%%%%%%%%%%%%%%%%%%%%%%%%%%%%%%%%%%%%%%%%%%%%%%%%%%%%%%%%%%%%%
\begin{figure}
\begin{center}
\includegraphics[width=21pc]{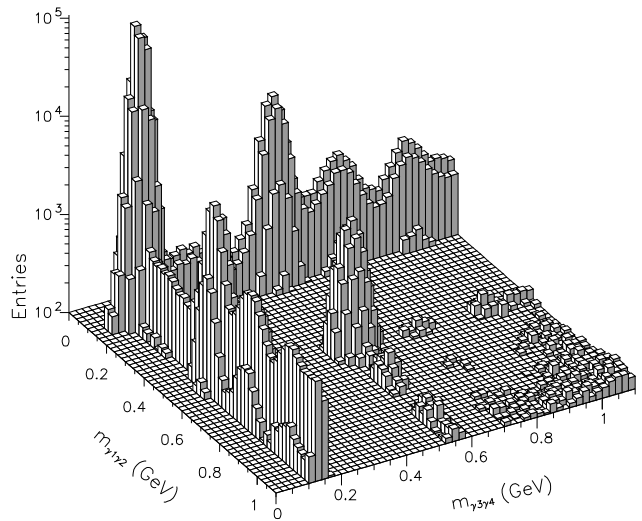}
\end{center}
\caption{\label{fig:2pspaper_3ent_scatterplot} 
LEGO plot of the two 2-photon invariant masses $m_{\gamma_1\gamma_2}$ and 
$m_{\gamma_3\gamma_4}$. Note the logarithmic scale of the vertical axis.}
\end{figure}
%%%%%%%%%%%%%%%%%%%%%%%%%%%%%%%%%%%%%%%%%%%%%%%%%%%%%%%%%%%%%%%%%%%%%%%%

Offline, events with four CCAL energy deposits greater than 100~MeV 
are retained if they passed a 5\% probability cut on a 
4-constraint (4C) fit to $\bar{p}p\rightarrow 4 \gamma$.
There are three ways to combine the four photons into two pairs; 
the event topology was
chosen as the combination with the highest confidence level of a 6C fit to
$\bar{p}p\rightarrow P_{1}P_{2} \rightarrow 4 \gamma$. 
However, given the limited opening angles of the decay
$\pi^0 (\mathrm{or}~\eta)\rightarrow\gamma\gamma$ at these energies
and the large angle between the two mesons, it is virtually impossible 
that more than one combination satisfy the 6C fits for more than one of the processes~(1)-(3).
Figure~\ref{fig:2pspaper_3ent_scatterplot} is a LEGO plot of 
the two 2-photon invariant masses prior to any fitting.
Evident are  high peaks of $\pi^0\pi^0$, $\pi^0\eta$ 
and $\eta\eta$ events, as well as those of $\pi^0\omega$ 
(where a photon in the $\omega\rightarrow\pi^0\gamma\rightarrow\gamma\gamma\gamma$ 
decay chain is not observed), $\pi^0\eta'$ and $\eta\eta'$. 
Marked ``berms'' are produced by $\pi^0 X$ events, where $X$
can be one or more particles that partially escaped detection.
Softer $\eta X$ berms are also noticeable.
Prior to a 4C fit, the mass resolution for a $\pi^0$, an $\eta$, and an $\eta'$ is 
$\sigma$$\simeq$$15$~MeV, $25$~MeV and $35$~MeV, respectively, 
with a small dependence on $z$.
\begin{table*}
\caption{\label{table-kincuts}The kinematic cuts applied to each channel.
The asymmetry of the cut on the invariant
mass reconstructed of the candidate $\pi^0$ 
compensates for a slight overestimate of the invariant mass reconstructed
for  $\pi^0$'s when the energy deposits from the daughter photons overlap 
in the CCAL (see details in \cite{bigpaper}).}
\begin{center}
\begin{tabular}{|c|c|c|c|c|}\hline
{Channel} & $\vert\Delta\theta\vert$ & 
$\vert\Delta\phi\vert$ & $m_{\gamma_1\gamma_2}$~cut  & $m_{\gamma_3\gamma_4}$~cut  \\
\hline
$\pi^{0}\,\pi^{0}$ & 12~mrad & 30~mrad & $100$~MeV~-~$185$~MeV 
 & $100$~MeV~-~$185$~MeV \\
$\pi^{0}\,\eta~   $ & 14~mrad & 38~mrad & $100$~MeV~-~$185$~MeV 
 & $M_{\eta}\pm 70$~MeV\\
$\eta\,\eta~~    $ & 15~mrad & 45~mrad & $M_{\eta}\pm 70$~MeV  
 & $M_{\eta}\pm 70$~MeV\\
$\pi^{0}\,\eta'  $ & 18~mrad & 50~mrad & $100$~MeV~-~$185$~MeV 
 & $M_{\eta'}\pm 40$~MeV\\
$\eta\,\eta'~   $ & 18~mrad & 50~mrad & $M_{\eta}\pm 45$~MeV  
 & $M_{\eta'}\pm 40$~MeV\\
\hline
\end{tabular}
\end{center}
\end{table*}
 
A set of kinematic cuts is applied for each process with the
values shown in Table~\ref{table-kincuts}. These include 
cuts on the invariant masses of each pair of photons
and on the colinearity  and coplanarity of the candidate mesons. 
The colinearity, $\Delta\theta$, of the two mesons~A and~B, 
is given by $\theta_{\mathrm{meas}}^\mathrm{A}-\theta_{\mathrm{calc}}^\mathrm{A}$,
where $\theta_{\mathrm{meas}}^\mathrm{A}$ is the {\it{measured}} polar angle of 
meson~A while $\theta_{\mathrm{calc}}^\mathrm{A}$ is the value of 
the same quantity {\it{calculated}} assuming two-body kinematics
using the measured polar angle of meson~B.
In the case where the two mesons are the same, A is chosen
to be the one in the forward direction.
In the case where they are different, A is chosen to be the 
lighter of the two mesons. 
The coplanarity, $\Delta\phi$, is defined as 
$180^\circ-\vert\phi_{\mathrm{meas}}^\mathrm{A}-
                \phi_{\mathrm{meas}}^\mathrm{B}\vert$, where 
$\phi_{\mathrm{meas}}^\mathrm{A}$ and 
$\phi_{\mathrm{meas}}^\mathrm{B}$ are the measured azimuthal angles 
of the 
mesons.
%
%%%%%%%%%%%%%%%%%%%%%%%%%%%%%%%%%%%%%%%%%%%%%%%%%%%%%%%%%%%%%%%%%%%%%%%%
\begin{figure*}
\begin{center}
\includegraphics[width=35pc]{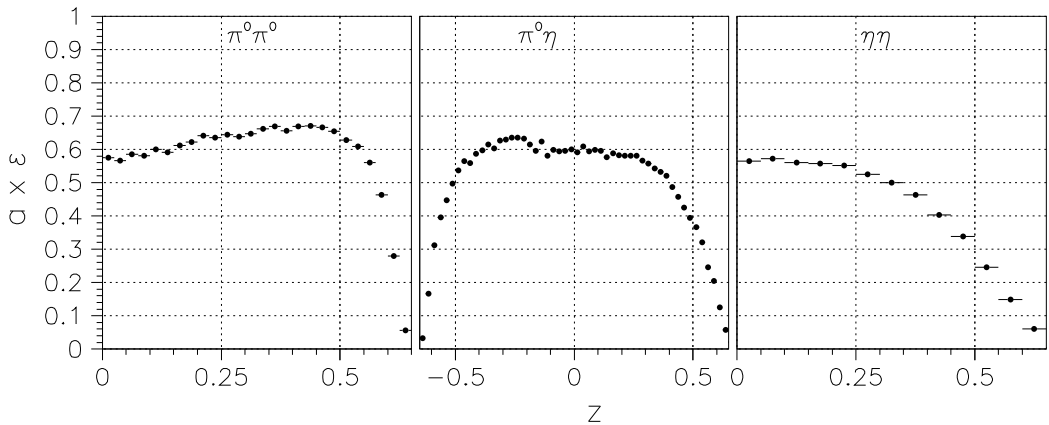}
\end{center}
\caption{         \label{fig:2pspaper_eff_costh} 
The product of detector acceptance and selection efficiency,
$a$$\times$$\epsilon$, as a function of $z$ for $\pi^0\pi^0$, $\pi^0\eta$ and
$\eta\eta$ events.
In the $\pi^0\eta$ case, where the two mesons are distinguishable, $z$ is defined
as $\cos\theta^*_{\pi^0}$.}
\end{figure*}
%%%%%%%%%%%%%%%%%%%%%%%%%%%%%%%%%%%%%%%%%%%%%%%%%%%%%%%%%%%%%%%%%%%%%%%%

The selection efficiency ($\epsilon$) and detector acceptance ($a$) 
are determined by Monte Carlo (MC) simulation. 
MC events are generated with a uniform angular distribution and a matrix is
built to record both the generated and reconstructed values of $z$.
The angular distribution of the  data is determined by using the 
inverse of this matrix to correct for effects due to the angular 
resolution of the detector.
The product of the acceptance and efficiency ($a$$\times$$\epsilon$) 
as a function of the generated value of $z$ for $\pi^0\pi^0$, $\pi^0\eta$ and 
$\eta\eta$ is shown in Figure~\ref{fig:2pspaper_eff_costh}. 

A significant source of inefficiency is  event pileup.
It is particularly important to correct for this 
effect since it is rate dependent and the instantaneous
luminosity varied from one energy point to another.
To determine the pileup effects, MC events were 
overlaid with the contents of random-gate events recorded throughout
the data-taking, thus reproducing the conditions of each energy point.
The event reconstruction was then performed on these hybrid events to
determine the reconstruction efficiency.
The rate-dependent losses vary from 14\% to 23\% with an average of 20\%; 
these losses do not differ significantly among the analyzed reactions.
As a function of $z$, the pileup correction is determined and applied.

\section{\label{sec:background}Signal and Background Subtraction 
for \bm{$\pi^{0}\pi^{0}$}, \bm{$\pi^{0}\eta$} and \bm{$\eta\eta$} samples.}

\begin{figure}
\begin{center}
\includegraphics[width=21pc]{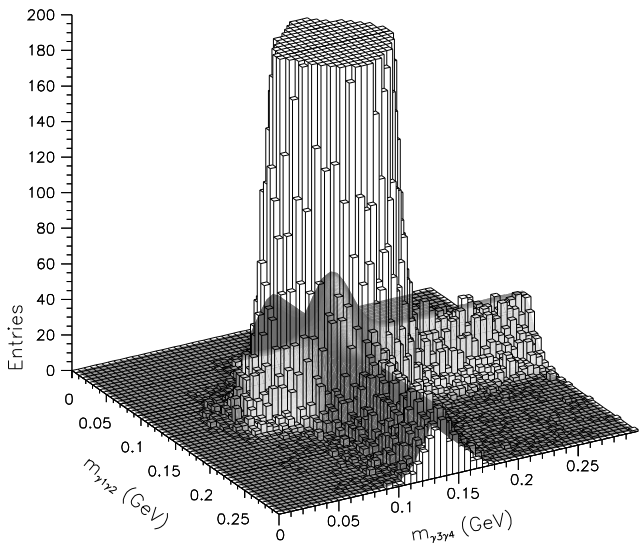}
\end{center}
\caption{         \label{fig:2pspaper_2pi0_overlap} 
The region of the $\pi^0\pi^0$ peak (truncated at about 2\% of its height) 
with a fit to the background shown as a gray surface.
}
\end{figure}

\begin{figure*}
\begin{center}
\includegraphics[width=35pc]{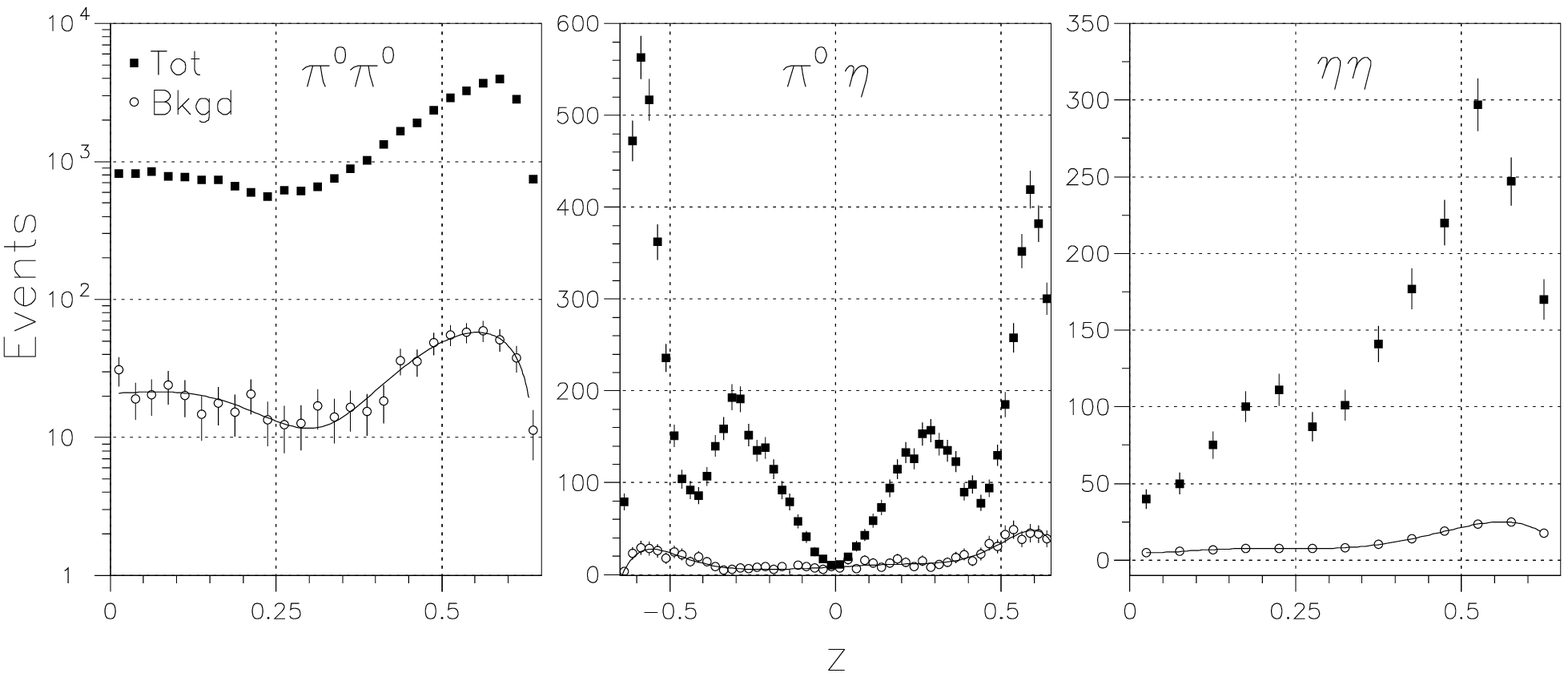}
\end{center}
\caption{         \label{fig:2pspaper_bkg_costh} 
The number of candidate events (Tot) as a function of $z$ at the $\chi_{c0}$ mass,
the estimated background events (Bkgd), and a polynomial fit (solid curve) 
to the number of background events.
Corrections for detector acceptance and efficiency have not yet been applied.
For the $\eta\eta$ channel, the background distribution was independent of 
energy and the data from all 17 energy points have been merged to 
reduce bin-to-bin fluctuations.
}
\end{figure*}

Figure~\ref{fig:2pspaper_2pi0_overlap} 
shows the region of the $\pi^0\pi^0$ peak 
before applying the mass cuts. 
A log-likelihood fit is performed to this data. 
The function used is the sum of a 
two gaussians, having the same mean,
to describe the $\pi^0\pi^0$ 
peak, and two gaussian berms and a tilted plane to describe the background
(from events such as $\pi^0\omega\rightarrow\pi^0\pi^0\gamma$
and  $\pi^0\pi^0\pi^0$ where, respectively, one and two photons are not observed).
The background component of the fitted function is shown as a gray surface. 
The background is estimated and subtracted as a function of $z$ as 
indicated in Figure~\ref{fig:2pspaper_bkg_costh}
and amounts to ($2-2.5$)\% for $z<0.3$ and ($1.5-2$)\% for $0.3<z<0.6$.
%
%%%%%%%%%%%%%%%%%%%%%%%%%%%%%%%%%%%%%%%%%%%%%%%%%%%%%%%%%%%%%%%%%%%%%%%%
%%%%%%%%%%%%%%%%%%%%%%%%%%%%%%%%%%%%%%%%%%%%%%%%%%%%%%%%%%%%%%%%%%%%%%%%

\begin{figure}
\begin{center}
\includegraphics[width=21pc]{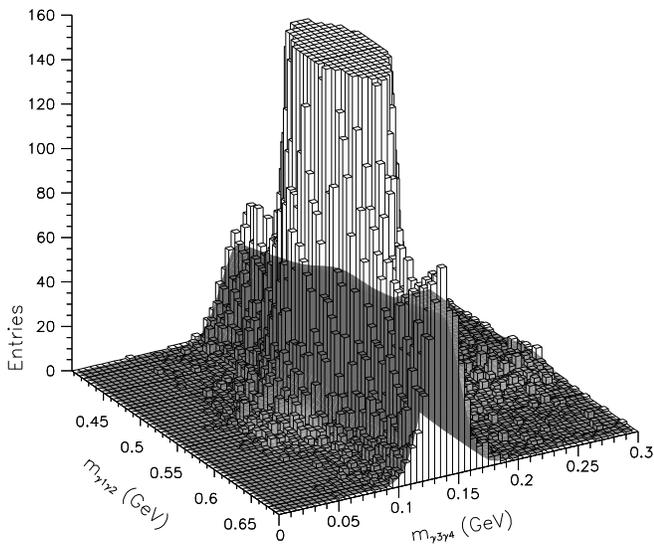}
\end{center}
\caption{         \label{fig:2pspaper_pi0eta_overlap} 
The region of the $\pi^0\eta$ peak (truncated at about 20\% of its height) 
with a fit to the background shown as a gray surface.
}
\end{figure}

A similar procedure is used in Figure~\ref{fig:2pspaper_pi0eta_overlap}
for determining the background at the $\pi^0\eta$ peak, which amounts to a 
total of $11\%$ over the range $-0.6<z<0.6$ and is distributed as
shown in Figure~\ref{fig:2pspaper_bkg_costh}.
%
%%%%%%%%%%%%%%%%%%%%%%%%%%%%%%%%%%%%%%%%%%%%%%%%%%%%%%%%%%%%%%%%%%%%%%%%
%%%%%%%%%%%%%%%%%%%%%%%%%%%%%%%%%%%%%%%%%%%%%%%%%%%%%%%%%%%%%%%%%%%%%%%%

\begin{figure}
\begin{center}
\includegraphics[width=21pc]{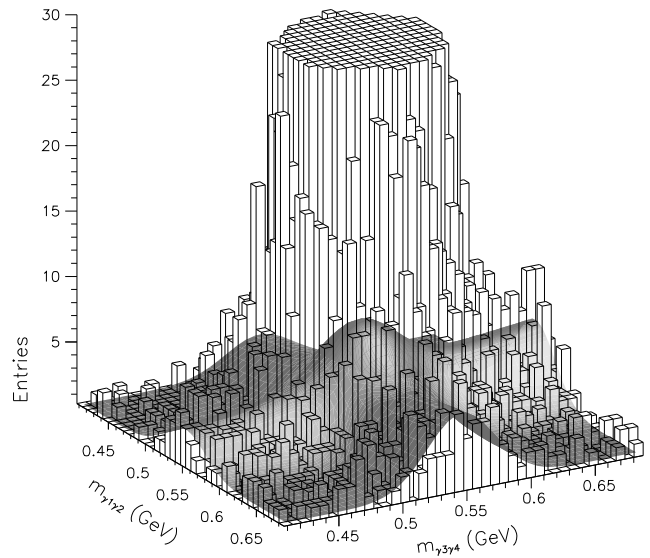}
\end{center}
\caption{         \label{fig:2pspaper_2eta_overlap} 
The region of the $\eta\eta$ peak (truncated at about 12\% of its height) 
with a fit to the background shown as a gray surface.
}
\end{figure}

Figure~\ref{fig:2pspaper_2eta_overlap} shows the $\eta\eta$ peak and its
fitted background, amounting to a total of $8\%$ over the range $0<z<0.6$
with the angular distribution shown in Figure~\ref{fig:2pspaper_bkg_costh}.
%
%%%%%%%%%%%%%%%%%%%%%%%%%%%%%%%%%%%%%%%%%%%%%%%%%%%%%%%%%%%%%%%%%%%%%%%%
%%%%%%%%%%%%%%%%%%%%%%%%%%%%%%%%%%%%%%%%%%%%%%%%%%%%%%%%%%%%%%%%%%%%%%%%

%%%%%%%%%%%%%%%%%%%%%%%%%%%%%%%%%%%%%%%%%%%%%%%%%%%%%%%%%%%%%%%%%%%%%%%%
%%%%%%%%%%%%%%%%%%%%%%%%%%%%%%%%%%%%%%%%%%%%%%%%%%%%%%%%%%%%%%%%%%%%%%%%

\section{\label{sec:2pi0_cross_section}The \bm{$\pi^0\pi^0$} Cross Section}

\begin{figure*}
\begin{center}
\includegraphics[width=35pc]{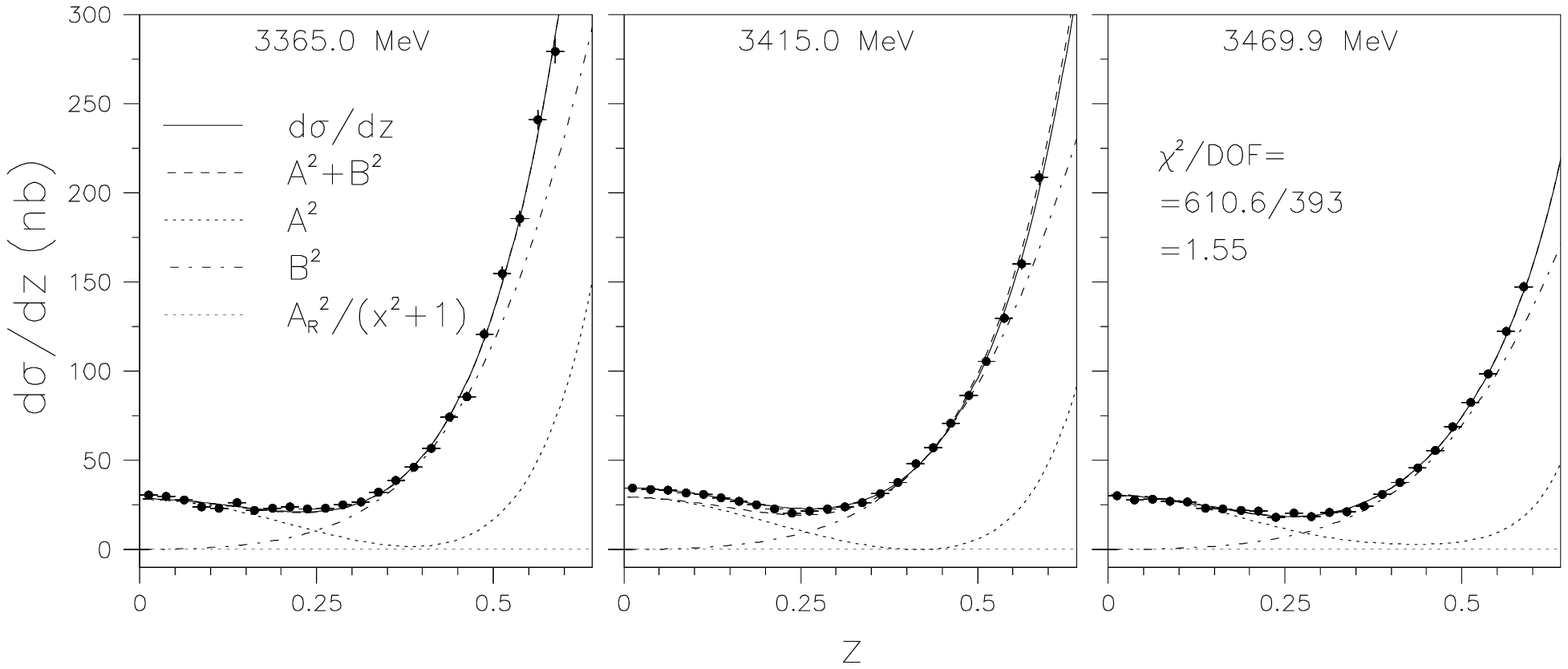}
\end{center}
\caption{         \label{fig:2pspaper_2pi0_diffcsfit_3stacks} 
The $\bar{p}p\rightarrow\pi^0\pi^0$ differential cross section versus 
$z\equiv\vert\cos\theta^*\vert$ at different $E_{cm}$. 
A fit using 
Equations~(\ref{eq:withampres_AandB})-(\ref{eq:noninterferingamp}), with 
values from  Table~\ref{tab:coefficients_alt}, is shown.
}
\end{figure*}

\begin{figure}
\begin{center}
\includegraphics[width=21pc]{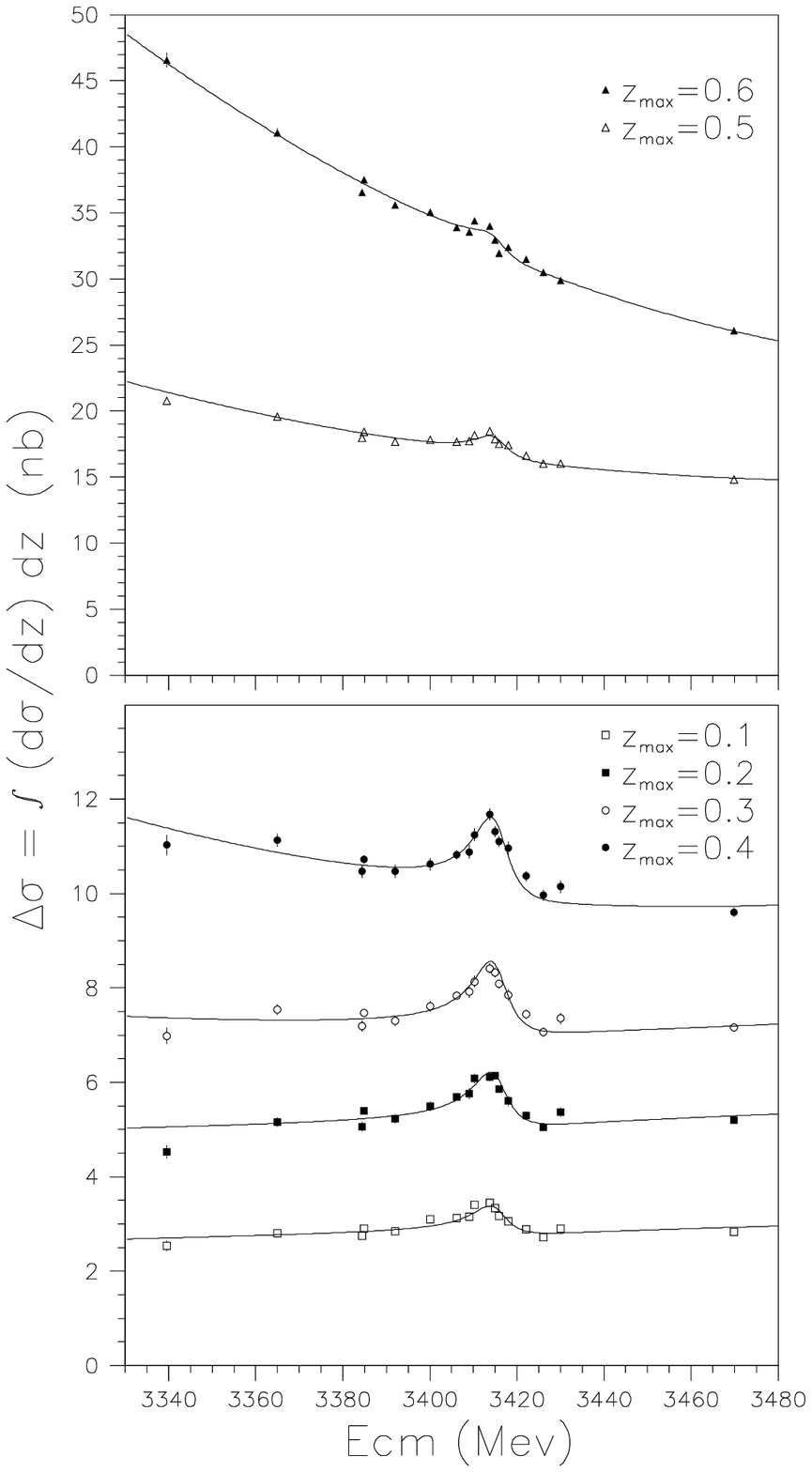}
\end{center}
\caption{         \label{fig:2pspaper_2pi0_csecm_fit_0_zmax} 
The measured $\bar{p}p\rightarrow\pi^0\pi^0$ cross section integrated over 
$0<z<z_{max}$ plotted versus $E_{cm}$.
A fit using 
Equations~(\ref{eq:withampres_AandB})-(\ref{eq:noninterferingamp}) is shown.
}
\end{figure}

A paper on the $\pi^0\pi^0$ reaction has been published~\cite{2pi0}.
The measured $\bar{p}p\rightarrow\pi^0\pi^0$ differential cross section 
is shown in Figure~\ref{fig:2pspaper_2pi0_diffcsfit_3stacks} for 3 of the 
17 energy points of the data sample and over an angular range limited
to $0<z<0.6$ by the detector acceptance. 
The cross section integrated to various values of $z_{max}$ is shown in 
Figure~\ref{fig:2pspaper_2pi0_csecm_fit_0_zmax} for all energy points.

%%%%%%%%%%%%%%%%%%%%%%%%%%%%%%%%%%%%%%%%%%%%%%%%%%%%%%%%%%%%%%%%%%%%%%%%
%%%%%%%%%%%%%%%%%%%%%%%%%%%%%%%%%%%%%%%%%%%%%%%%%%%%%%%%%%%%%%%%%%%%%%%%
%

%%%%%%%%%%%%%%%%%%%%%%%%%%%%%%%%%%%%%%%%%%%%%%%%%%%%%%%%%%%%%%%%%%%%%%%%
%%%%%%%%%%%%%%%%%%%%%%%%%%%%%%%%%%%%%%%%%%%%%%%%%%%%%%%%%%%%%%%%%%%%%%%%
Figures~\ref{fig:2pspaper_2pi0_diffcsfit_3stacks} and
\ref{fig:2pspaper_2pi0_csecm_fit_0_zmax} also show a
binned maximum-likelihood fit to the cross section. 
The fit is performed simultaneously on all 17 energy points 
and over $0<z<0.6$. Within this range, the number of 
background-subtracted $\pi^0\pi^0$ events is 431,625.
The parameterization of 
Equations~(\ref{eq:withampres_AandB})-(\ref{eq:noninterferingamp}) 
is used setting $J_{max}=4$;
the number of partial waves to include and their energy dependence
were determined by searching for significant improvements of the $\chi^2$.
The mass and width of the $\chi_{c0}$ are constrained to the values 
(see Table~\ref{tab:e835_results1}) determined by studying the process
$\bar{p}p\rightarrow J/\psi\,\gamma;J/\psi\rightarrow e^+e^-$ \cite{psigamma}.
A magnified version of the plot of $d\sigma/dz$ at $E_{cm}=3415$~MeV, 
as well as a detailed description of the fit and its different components,
is provided in \cite{2pi0}.
The fit results are given in Table~\ref{tab:coefficients_alt}.

\begin{table*}
\caption{\label{tab:coefficients_alt}
Fit results for coefficients and phases of the partial-wave expansion of 
Equations~(\ref{eq:withampres_AandB})-(\ref{eq:noninterferingamp})
for the $\pi^0\pi^0$, $\pi^0\eta$ and $\eta\eta$ channels. 
A linear energy dependence is included when found necessary.
The errors are statistical.
}
\begin{tabular}{|c|cc|}\hline
            &~$C_J~[~\mathrm{nb}^{1/2}~]~\&~C_J^1~[~\mathrm{nb}^{1/2}~]$
            &~$\delta_J~[~\mathrm{degree}~]~\&~\delta_J^1~[~\mathrm{degree}~]  $\\
\hline
            &~$C_0=(12.8\pm0.7)-(0.19\pm0.02) x$
            &~$\delta_0=-36.1\pm1.8                                           $\\
            &~$C_2=(8.3\pm0.4)-(0.055\pm0.010) x$
            &~$\delta_2=-43.0\pm1.0                                           $\\
~$\pi^0\pi^0$~&~$C_4=(2.5\pm0.2)+(0.022\pm0.003) x$
            &~$\delta_4=-15.1\pm1.2                                           $\\ 
            &~$C_2^1=(5.19\pm0.13)-(0.063\pm0.005) x$
            &~$-                                                             $\\
            &~$C_4^1=(1.52\pm0.09)-(0.021\pm0.003) x$
            &~$\delta_4^1-\delta_2^1=-0.7\pm0.2                               $\\ 
\hline
            &~$C_0=(19.7\pm0.3)-(0.076\pm0.003) x$
            &~$-                                                             $\\
            &~$C_2=(12.2\pm0.2)$
            &~$\delta_2-\delta_0=4.5\pm0.8                                    $\\
~$\pi^0\eta$~&~$C_4=(3.21\pm0.09)$
            &~$\delta_4-\delta_0=9\pm2                                        $\\ 
            &~$C_2^1=(2.8\pm0.2)-(0.129\pm0.008) x$
            &~$-                                                             $\\
            &~$C_4^1=(2.54\pm0.09)-(0.087\pm0.004) x$
            &~$\delta_4^1-\delta_2^1=0                                        $\\ 
\hline
            &~$C_0=(7.27\pm0.11)-(0.269\pm0.012) x$
            &~$\delta_0=138.2\pm2.9                                           $\\
            &~$C_2=(2.07\pm0.06)-(0.116\pm0.006) x$
            &~$\delta_2=138.2\pm2.9                                           $\\
~$\eta\eta $~&~$C_4=0$
            &~$\delta_4=138.2\pm2.9                                           $\\
            &~$C_2^1=(4.56\pm0.07) x$
            &~$-                                                             $\\
            &~$C_4^1=(2.65\pm0.05) x$
            &~$\delta_4^1-\delta_2^1=0                                        $\\
\hline
\end{tabular}
\end{table*}

The fit presented in this section demonstrates the 
general structure of the $\bar{p}p\rightarrow\pi^0\pi^0$ angular distribution 
and estimates the number and amount of the contributing partial waves.
However, this fit is not very sensitive to the value of the 
resonant amplitude $A_R$.
The reason is that the size of the (interference-enhanced) resonant signal
is significant only at small values of $z$, 
while the fit is dominated by the high statistics of the 
nonresonant forward peak.
The next section describes how the extraction of $A_R$ is carried out.

\section{\label{sec:2pi0_ampres}Extraction of
\bm{$B(\chi_{c0}\rightarrow\bar{p}p)\times B(\chi_{c0}\rightarrow\pi^0\pi^0)$} }

As discussed in Section~\ref{sec:intro}
(see in particular the discussion of Equation~\ref{eq:interfterm}),
the natural region to exploit in order to obtain a model-insensitive
measurement of 
the resonance amplitude $A_R$ is the region at small values of~$z$.

To find the value of $z$ at which the noninterfering continuum starts to
play a significant role, we perform  independent 
fits in each bin $\Delta z$ at small $z$.
Each fit uses the parameterization of Equation~(\ref{eq:withampres_AandB}) with 
$Be^{i\delta_B}$ set equal to zero.
The parameterization used for the helicity-0 component of the continuum is 
%%%%%%%%%%%%%%%%%%%%%%%%%%%%%%%%%%%%%%%%%%%%%%%%%%%%%%%%%%%%%%%%%%%%%%%%%%%%%
\begin{equation}
A^2\equiv a_0 + a_1 x + a_2 x^2\,,  
\label{eq:2pi0_indep_A_Bequal0}
\end{equation}
%%%%%%%%%%%%%%%%%%%%%%%%%%%%%%%%%%%%%%%%%%%%%%%%%%%%%%%%%%%%%%%%%%%%%%%%%%%%%
to reproduce the moderate energy dependence  
of the cross section, see Figure~\ref{fig:2pspaper_2pi0_csecm_fit_0_zmax}.
 
\begin{figure}
\begin{center}
\includegraphics[width=21pc]{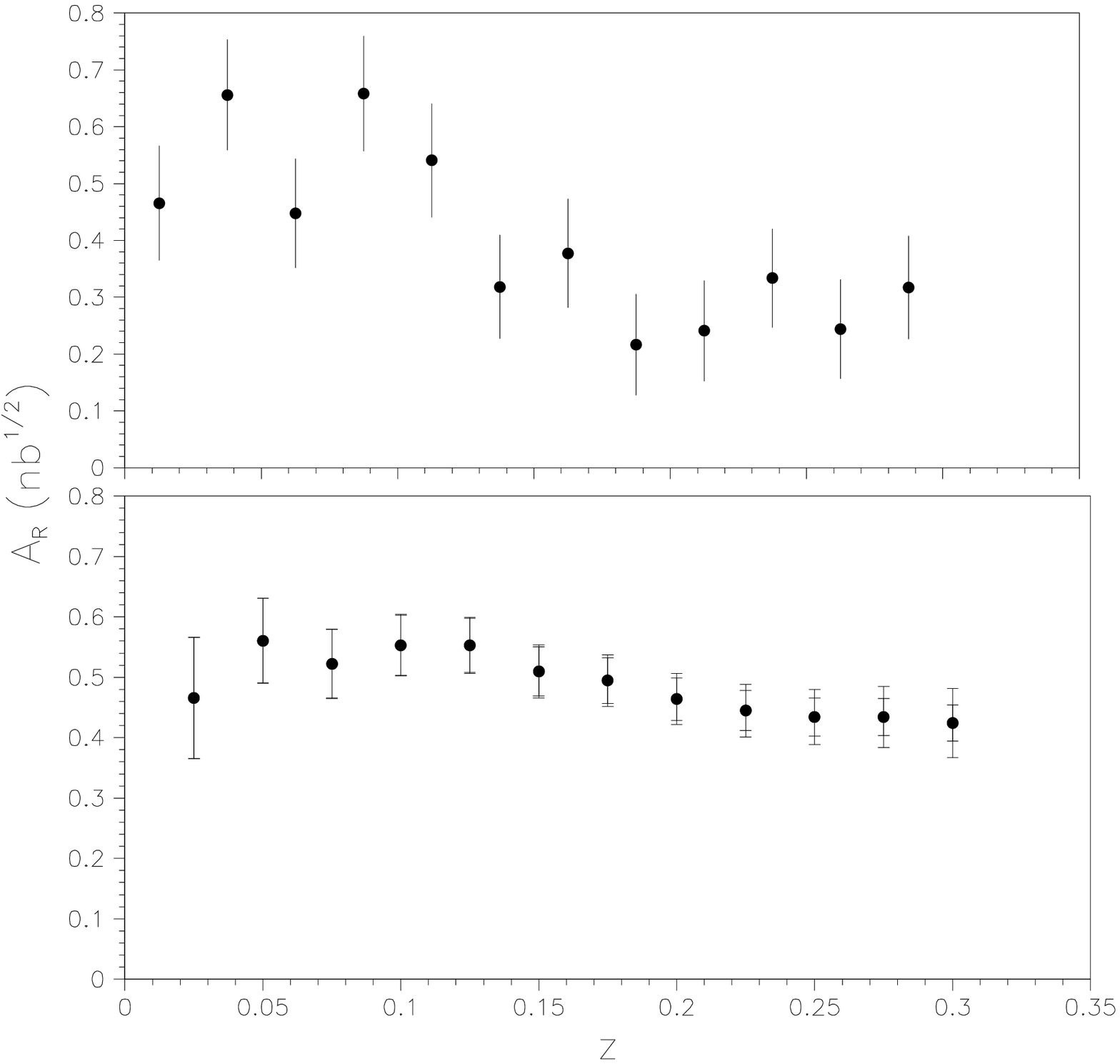}
\end{center}
\caption{         \label{fig:2pi0_ampres} 
Fits for the resonant amplitude 
of $\bar{p}p\rightarrow\chi_{c0}\rightarrow\pi^0\pi^0$. 
Top: the fits are to the data of each bin in $\Delta z=0.025$. 
$B$ is set to zero (see text). 
Bottom: each fit is performed over a range $0<z<z_{max}$ for increasing $z_{max}$ 
and $\vert B~e^{i\delta_B} \vert^2$ is inserted from the  
fit of Figure~\ref{fig:2pspaper_2pi0_diffcsfit_3stacks}.
(The inner error bars are statistical while the outer 
ones indicate the uncertainty on $C_2^1$, $C_4^1$ and $\delta_4^1$-$\delta_2^1$ 
in Table~\ref{tab:coefficients_alt}).
}
\end{figure}

Since the growth with $z$ of the helicity-1 component is not accounted
for in these fits, the estimate of $A_R$, which in reality is a 
constant, will show an artificial decrease at values of z where the 
helicity-1 component can
no longer be neglected. The value of $A_{R}$ can then be derived 
by using all the data up to some maximum value of $z$ identified by the
fits to the individual bins. 
The fit results for $A_R$  from individual bins of $\Delta z$ up to $z=0.3$ 
are shown in Figure~\ref{fig:2pi0_ampres}-top and the values of $A_{R}$
derived by using the data within $0<z<z_{max}$ are shown in
Figure~\ref{fig:2pi0_ampres}-bottom.

\begin{figure}
\begin{center}
\includegraphics[width=21pc]{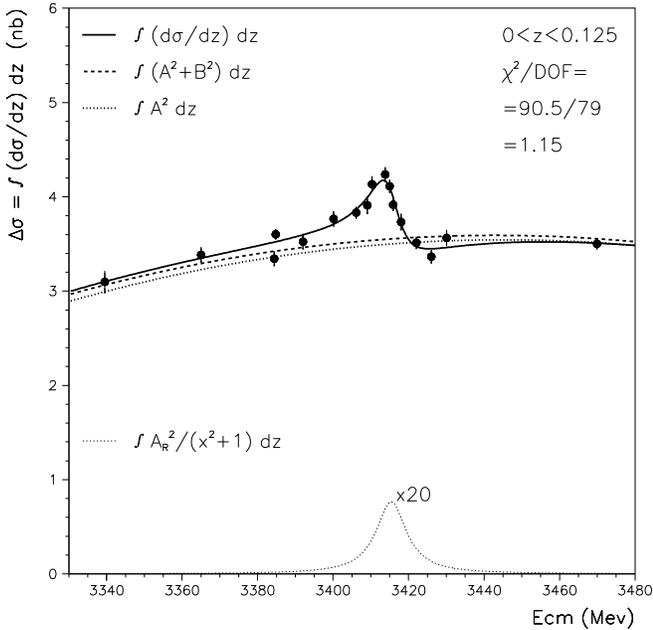}
\end{center}
\caption{         \label{fig:2pi0_csecmfit_bin1_5} 
The $\bar{p}p\rightarrow\pi^0\pi^0$ cross section integrated over 
$0<z<0.125$ plotted versus $E_{cm}$.
The fit of Equation~\ref{eq:withampres_AandB} is shown along with its components.
The contribution of $\vert B~e^{i\delta_B} \vert^2$ is 
constrained to the estimate obtained from the partial-wave expansion fit of 
Figure~\ref{fig:2pspaper_2pi0_diffcsfit_3stacks}, and its integration
corresponds to the separation between the 
curves $\int\,(A^2+B^2)\,dz$ and $\int\,A^2\,dz$. 
The ``pure'' Breit-Wigner curve, $\int [A_R^2/(x^2+1)]\,dz$, 
is shown multiplied by a factor 20.
}
\end{figure}

The observed drop in $A_{R}$ of Figure~\ref{fig:2pi0_ampres}-top 
at $z \gtrsim 0.15$ 
shows that the helicity-1 component is no longer negligible at this value 
of $z$. We therefore restrict our region to 
\begin{equation}
z\le z_{max}\equiv 0.125
\label{eq:2pi0_zcut}
\end{equation}
The restricted $z$ range increases the statistical uncertainty on $A_R$, but 
ensures there is little effect on $A_R$ from systematic uncertainties in 
our knowledge of $B$.
The fit is described in detail in \cite{2pi0} and is reproduced 
in Figure~\ref{fig:2pi0_csecmfit_bin1_5}.
%%%%%%%%%%%%%%%%%%%%%%%%%%%%%%%%%%%%%%%%%%%%%%%%%%%%%%%%%%%%%%%%%%%
%%%%%%%%%%%%%%%%%%%%%%%%%%%%%%%%%%%%%%%%%%%%%%%%%%%%%%%%%%%%%%%%%%%
%%%%%%%%%%%%%%%%%%%%%%%%%%%%%%%%%%%%%%%%%%%%%%%%%%%%%%%%%%%%%%%%%%%%%%%%
%%%%%%%%%%%%%%%%%%%%%%%%%%%%%%%%%%%%%%%%%%%%%%%%%%%%%%%%%%%%%%%%%%%%%%%%

The value obtained for the resonant amplitude $A_R$ is given by:
\begin{equation}
A_R^2= \pi \lambdabar^2 \times
B(\chi_{c0} \rightarrow \bar{p}p)\times B(\chi_{c0} \rightarrow\pi^0\pi^0)~,
\label{eq:2pi0_sigmaR_BinBout}
\end{equation}
where $\lambdabar$ is the center-of-mass de Broglie wavelength, which gives:
\begin{equation}
B(\chi_{c0} \rightarrow \bar{p}p)\times B(\chi_{c0} \rightarrow\pi^0\pi^0)=
(5.09\pm0.81\pm0.25)\times 10^{-7}~.
\label{eq:2pi0_BinBout}
\end{equation}
The uncertainties are statistical and systematic, respectively.
The dominant systematic errors arise from the luminosity determination 
(2.5\%) and the knowledge of $M_{\chi_{c0}}$ from \cite{psigamma} (2.5\%). 
The systematic error due to the uncertainty in the helicity-1 continuum 
$\vert B~e^{i\delta_B} \vert^2$ is $1.2\%$.

The phase between the 
helicity-0 nonresonant amplitude $A$ and the resonant amplitude $A_R$ is 
$\delta_A=(39\pm5\pm6)~\mathrm{degrees}$.
In the region $0<z<0.125$  of the fit, 
no dependence of $\delta_A$ on $z$ and $x$ is found.
The values of $\delta_0$, $\delta_2$, and $\delta_4$ are given in Table~\ref{tab:coefficients_alt}.

\section{\label{sec:2eta_cross_section}The \bm{$\eta\eta$} Cross Section}

\begin{figure*}
\begin{center}
\includegraphics[width=35pc]{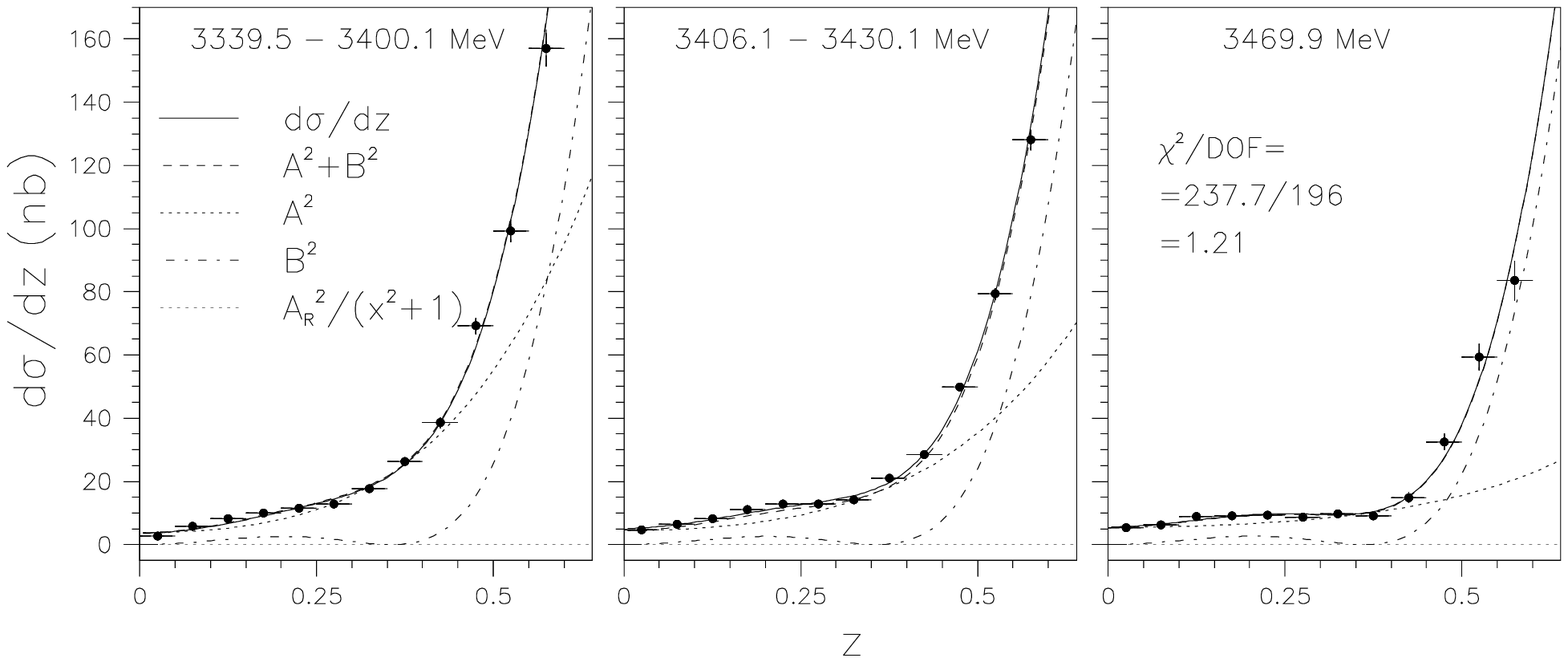}
\end{center}
\caption{         \label{fig:2pspaper_2eta_diffcsfit_3groups} 
The $\bar{p}p\rightarrow\eta\eta$ differential cross section versus 
$z\equiv\vert\cos\theta^*\vert$ at different $E_{cm}$. 
To reduce statistical fluctuations, the 17 energies are displayed merged as
$3339.5$~MeV~$<E_{cm}<3400.1$~MeV (left), $3406.1$~MeV~$<E_{cm}<3430.1~$MeV 
(center) and $E_{cm}=3469.9$~MeV (right).
A fit using Equations~(\ref{eq:withampres_AandB})-(\ref{eq:noninterferingamp})
is shown along with its components (see Table~\ref{tab:coefficients_alt} 
for values of fit parameters).
}
\end{figure*}

\begin{figure}
\begin{center}
\includegraphics[width=21pc]{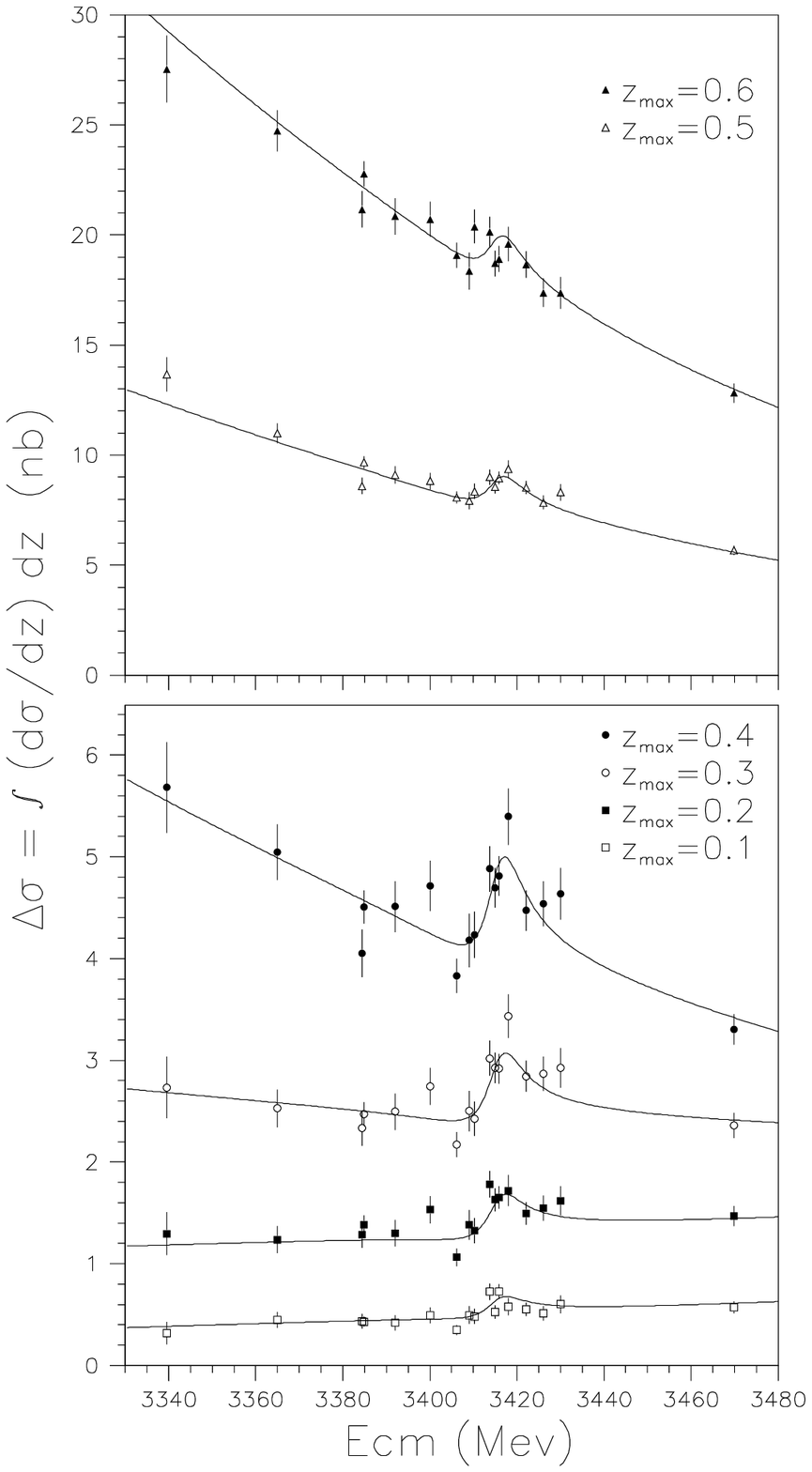}
\end{center}
\caption{         \label{fig:2pspaper_2eta_csecm_fit_0_zmax} 
The measured $\bar{p}p\rightarrow\eta\eta$ cross section integrated over 
$0<z<z_{max}$ plotted versus $E_{cm}$.
A fit using 
Equations~(\ref{eq:withampres_AandB})-(\ref{eq:noninterferingamp}) is shown.
}
\end{figure}

The measured $\bar{p}p\rightarrow \eta\eta$ differential
cross section is shown in Figure~\ref{fig:2pspaper_2eta_diffcsfit_3groups}.
Figure~\ref{fig:2pspaper_2eta_csecm_fit_0_zmax} shows the 
integrated cross section versus $E_{cm}$.
As in the $\pi^0\pi^0$ case, the cross section is dominated by the
nonresonant continuum $\bar{p}p\rightarrow\eta\eta$, 
which has a smooth dependence on the energy.
The $\eta\eta$ channel also shows a resonance signal near the  
mass of the $\chi_{c0}$.
The interference pattern is different from the one observed 
in the $\pi^0\pi^0$ channel. 
There is destructive interference on the low-energy side of the 
resonance and constructive on the high-energy side, with the resonance
peak shifted to above the $\chi_{c0}$ mass.

%%%%%%%%%%%%%%%%%%%%%%%%%%%%%%%%%%%%%%%%%%%%%%%%%%%%%%%%%%%%%%%%%%%%%%%%
%%%%%%%%%%%%%%%%%%%%%%%%%%%%%%%%%%%%%%%%%%%%%%%%%%%%%%%%%%%%%%%%%%%%%%%%
%%%%%%%%%%%%%%%%%%%%%%%%%%%%%%%%%%%%%%%%%%%%%%%%%%%%%%%%%%%%%%%%%%%%%%%%
%%%%%%%%%%%%%%%%%%%%%%%%%%%%%%%%%%%%%%%%%%%%%%%%%%%%%%%%%%%%%%%%%%%%%%%%
As in the $\pi^0\pi^0$ analysis,
a binned maximum-likelihood fit to the $\bar{p}p\rightarrow\eta\eta$
differential cross section is performed simultaneously on all 
17 energy points 
and over an angular range limited to $0<z<0.6$ by the detector acceptance.
Within this range, there are 19,675 background-subtracted $\eta\eta$ events.
The parameterization used is given in Equation~\ref{eq:withampres_AandB},
with the partial-wave expansion of Equations~\ref{eq:interferingamp} 
and \ref{eq:noninterferingamp}.
The mass and width of the $\chi_{c0}$ are constrained to the values 
determined by means of the $J/\psi\,\gamma$ channel \cite{psigamma}.

The result of the fit is shown in 
Figures~\ref{fig:2pspaper_2eta_diffcsfit_3groups}
and \ref{fig:2pspaper_2eta_csecm_fit_0_zmax}.
As in the $\pi^0\pi^0$ case, the contribution of the ``pure'' resonance, 
$A_R^2/(x^2+1)$, is negligible.
The $B^2$ (helicity-1) term dominates at larger $z$, but its growth 
begins at a larger value of $z$ than in the $\pi^0\pi^0$ case.
The interfering helicity-0 continuum, of magnitude $A^2$,
dominates for a large part of the angular range ($0<z\lesssim0.5$)
and provides the amplification for the interference-enhanced resonance 
signal seen in the separation between $d\sigma / dz$ and $A^2+B^2$.
Since the number of events available in the $\eta\eta$ channel is limited 
(about 20 times fewer than in the $\pi^0\pi^0$ channel),
the angular distribution within the available range in $z$ 
is not as well resolved as in the $\pi^0\pi^0$ and $\pi^0\eta$ channels. 
Compared to the $\pi^0\pi^0$ and $\pi^0\eta$, 
the $\eta\eta$ channel requires a smaller number of parameters.
In the best fit (see Table~\ref{tab:coefficients_alt}) 
$C_0$ and $C_2$ have a linear 
energy dependence; $C_4$ vanishes and is subsequently constrained to zero;
and $C_{21}$ and $C_{41}$ do not require an energy dependence. 
The phases $\delta_0$, $\delta_2$ and $\delta_4$ are equal to each other, 
different from zero (that is the phase of the resonant amplitude is 
different from the phases of the nonresonant amplitudes) 
and constant in energy.  
The phases of the helicity-1 continuum
(only the difference between $\delta_4^1$ and $\delta_2^1$ is measurable) 
also are not significantly different from each other and
do not exhibit an energy dependence.
Note that even in the $\pi^0\pi^0$ and $\pi^0\eta$ cases the phases of the
nonresonant amplitudes of a given helicity are very similar to each other.

\section{\label{sec:2eta_ampres}Extraction of
\bm{$B(\chi_{c0}\rightarrow\bar{p}p)\times B(\chi_{c0}\rightarrow\eta\eta)$} }

\begin{figure}
\begin{center}
\includegraphics[width=21pc]
{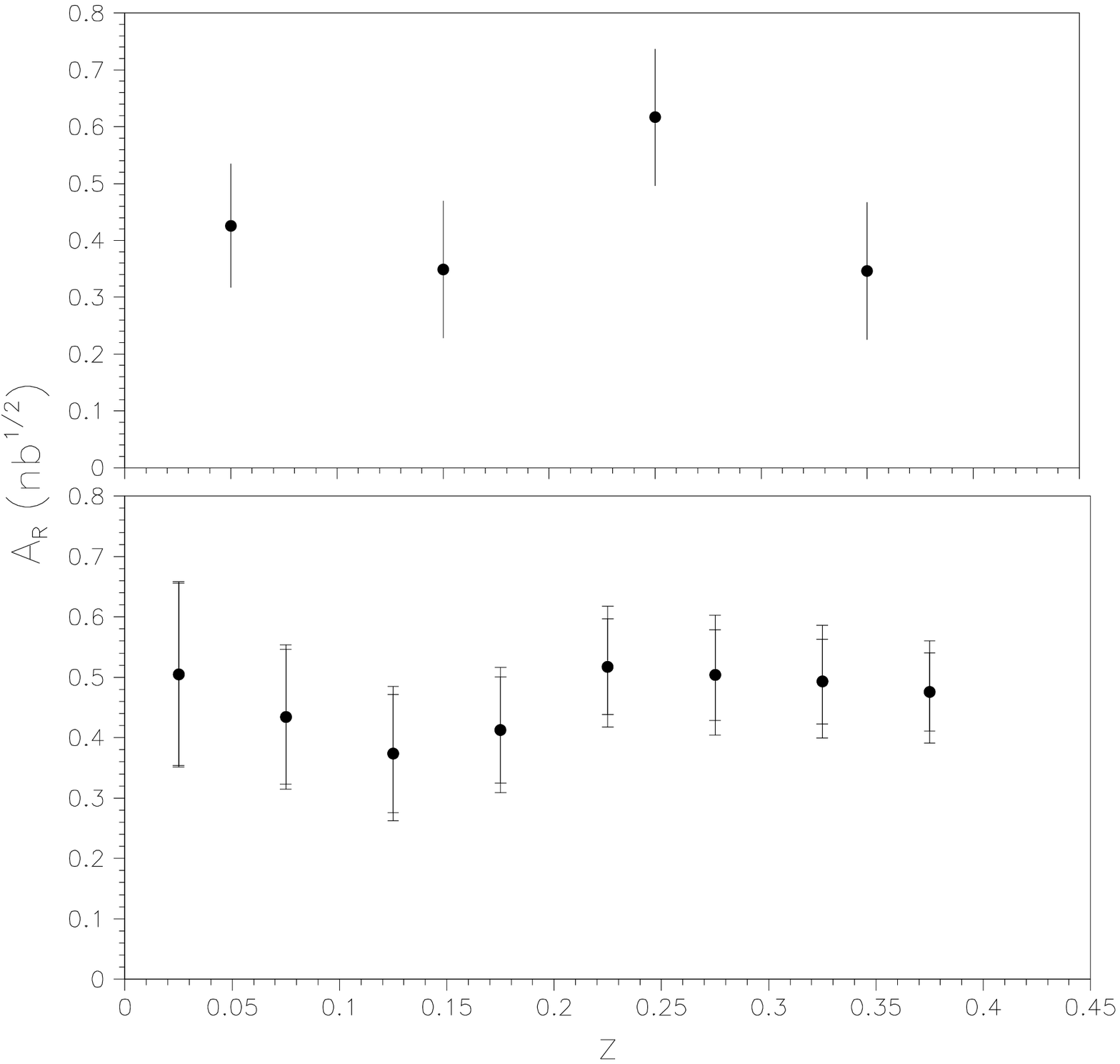}
\end{center}
\caption{         \label{fig:2eta_ampres} 
Fits for the resonant amplitude 
of $\bar{p}p\rightarrow\chi_{c0}\rightarrow\eta\eta$. 
Top: each fit is performed in a bin of size $\Delta z=0.1$ 
independently from the other bins; 
Equation~\ref{eq:withampres_AandB} is used with the helicity-1 
continuum $\vert B~e^{i\delta_B} \vert^2$ fixed to zero. 
Bottom: each fit is performed over a range $0<z<z_{max}$ for increasing $z_{max}$; 
Equations~\ref{eq:withampres_AandB} and \ref{eq:2eta_depend_A_zmax} are used, 
while $\vert B~e^{i\delta_B} \vert^2$ is taken from the partial-wave expansion 
fit of Figure~\ref{fig:2pspaper_2eta_diffcsfit_3groups} (the outer 
error bars show the uncertainty due to the errors on $C_2^1$ and $C_4^1$ in 
Table~\ref{tab:coefficients_alt}). 
}
\end{figure}

As in the $\pi^0\pi^0$ analysis, the effect of the helicity-1 component 
of the continuum is investigated by performing a series of independent 
fits on each bin $\Delta z$ with $Be^{i\delta_B}$ set equal to zero.
Figure~\ref{fig:2eta_ampres}-top shows the fit results for $A_R$ in every bin 
$\Delta z$ up to $z=0.4\,$.
%%%%%%%%%%%%%%%%%%%%%%%%%%%%%%%%%%%%%%%%%%%%%%%%%%%%%%%%%%%%%%%%%%%
%%%%%%%%%%%%%%%%%%%%%%%%%%%%%%%%%%%%%%%%%%%%%%%%%%%%%%%%%%%%%%%%%%%
No drop of $A_R$ is apparent in the region $0<z<0.4\,$, 
indicating that any significant rise of $B^2$ 
occurs at larger $z$ than in the $\pi^0\pi^0$ case. 
The fit in Figure~\ref{fig:2pspaper_2eta_diffcsfit_3groups}
also shows that $B^2$ is small compared to $A^2$ 
below $z\approx 0.4$\,. 
This is helpful because, in order to statistically resolve a clear resonant 
signal in the plot of the $\eta\eta$ cross section versus $E_{cm}$\,, 
it is necessary to integrate over a larger $z$-range 
than in the $\pi^0\pi^0$ analysis. 
Using a larger $z$-range produces a slightly larger uncertainty in the 
relative contributions of $A^2$ and $B^2$, and consequently in $A_R$. 

\begin{figure}
\begin{center}
\includegraphics[width=21pc]{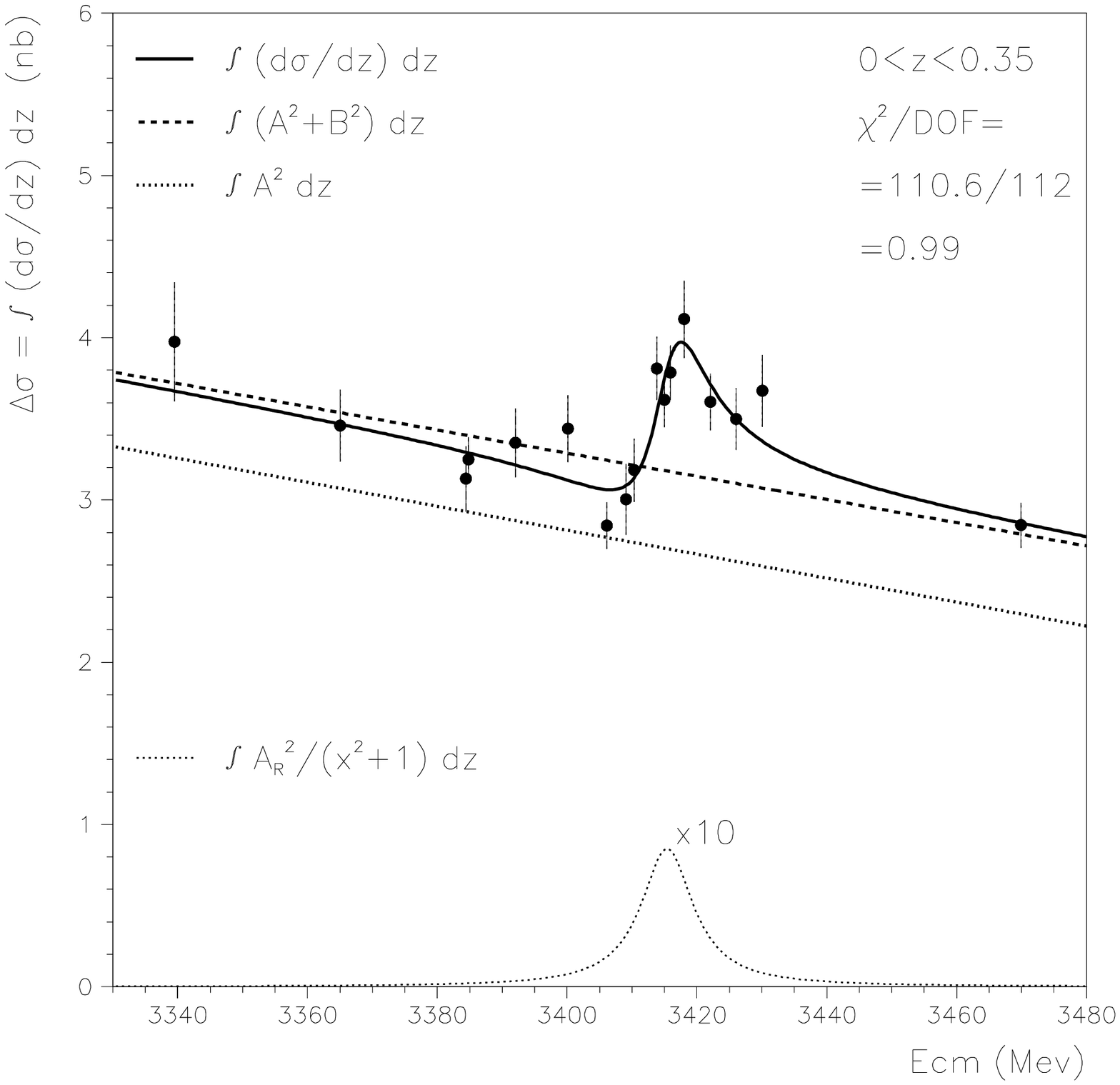}
\end{center}
\caption{         \label{fig:2eta_csecmfit_bin1_7} 
The $\bar{p}p\rightarrow\eta\eta$ cross section integrated over 
$0<z<0.35$ plotted versus $E_{cm}$.
The fit of Equation~\ref{eq:withampres_AandB}
and \ref{eq:2eta_depend_A_zmax} 
is shown along with its components.
The $\vert B~e^{i\delta_B} \vert^2$ term is
constrained to the estimate from the partial-wave expansion fit 
in Figure~\ref{fig:2pspaper_2eta_diffcsfit_3groups};
its contribution is shown by the separation
between the curves $~\int\,(A^2+B^2)\,dz$ and $\int\,A^2\,dz$. 
The ``pure'' Breit-Wigner curve, $\int A_R^2/(x^2+1)\,dz$, 
is shown multiplied by a factor 10 to make it comparable to the 
observed interference signal.
}
\end{figure}

As in the $\pi^0\pi^0$ analysis, the information from the different 
bins can be merged by performing new fits, integrating over ranges $0<z<z_{max}$ 
with increasing $z_{max}$.
Equation~\ref{eq:withampres_AandB} is used. 
The helicity-1 continuum is constrained to the estimate from the partial 
wave expansion fit of Figure~\ref{fig:2pspaper_2eta_diffcsfit_3groups}.
The parameterization of the magnitude of the helicity-0 continuum is
\begin{equation}
A^2\equiv a_0 + a_1 x + a_2 z^2 + a_3 z^2 x + a_4 z^4 ,
\label{eq:2eta_depend_A_zmax}
\end{equation}
where the necessary parameters are added, with increasing $z_{max}$, 
by searching for significant improvements in the $\chi^2$ of the fit.
The fact that an energy/polar-angle mixing term ($a_3 z^2 x$) 
is required is already noticeable in 
Figure~\ref{fig:2pspaper_2eta_csecm_fit_0_zmax}.
The phase $\delta_A$ does not require any significant dependence 
on $x$ or $z$.
The results for the resonant amplitude $A_R$ as a function of $z_{max}$
from these fits are shown in the bottom of
Figure~\ref{fig:2eta_ampres}.
The estimate of $A_R$ is stable for $z_{max}>0.2$\,and
the value of $A_R$ is extracted from the fit 
performed over the angular range $0<z<0.35\,$.
The result of this fit is shown in Figure~\ref{fig:2eta_csecmfit_bin1_7}.
The angular region of integration is larger than in the
$\pi^0\pi^0$ analysis, but still guarantees that the noninterfering 
helicity-1 continuum ($B^2$) is relatively small. 

%%%%%%%%%%%%%%%%%%%%%%%%%%%%%%%%%%%%%%%%%%%%%%%%%%%%%%%%%%%%%%%%%%%%%%%%

An equation analogous to Equation~\ref{eq:2pi0_sigmaR_BinBout} gives
\begin{equation}
B(\chi_{c0} \rightarrow \bar{p}p)\times B(\chi_{c0} \rightarrow\eta\eta)=
(4.0\pm1.2^{+0.5}_{-0.3})\times 10^{-7}~.
\label{eq:2eta_BinBout}
\end{equation}
The uncertainties are statistical and systematic, respectively.
The dominant systematic error arises from the uncertainty in the 
helicity-1 continuum $\vert B~e^{i\delta_B} \vert^2$: ($^{+9.6}_{-4.8}\%$).

The phase difference between the helicity-0 nonresonant amplitude $A$ and the 
resonant amplitude $A_R$ in the angular region $0<z<0.35$
is $\delta_A=(144\pm 8\pm 6)~\mathrm{degree}$;
no dependence on $x$ and $z$ is observed. 
The phase $\delta_A$, responsible for the shape of the interference pattern,
is different from the corresponding phase in the $\pi^0\pi^0$ channel.
In the $\eta\eta$ channel, the interference is destructive on the 
low-energy side of the resonance and constructive on the high-energy side.

\section{\label{sec:massgammafree} Estimate of \bm{$M_{\chi_{c0}}$} 
and \bm{$\Gamma_{\chi_{c0}}$} 
using the \bm{$\pi^0\pi^0$} and \bm{$\eta\eta$} samples alone} 

%%%%%%%%%%%%%%%%%%%%%%%%%%%%%%%%%%%%%%%%%%%%%%%%%%%%%%%%%%%%%%%%%%%%%%%%
\begin{figure}
\begin{center}
\includegraphics[width=21pc]{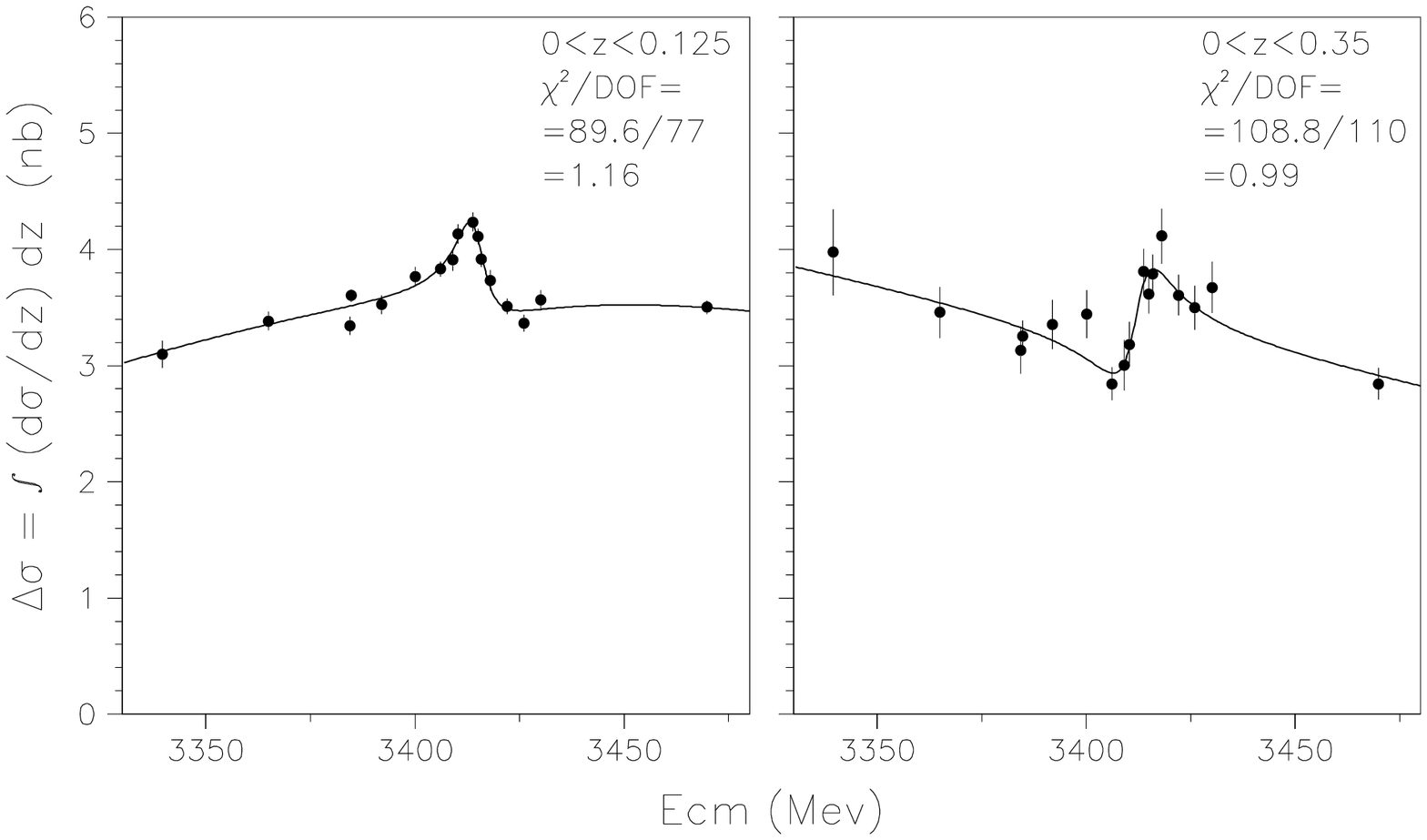}
\end{center}
\caption{         \label{fig:2pspaper_massgammafree} 
The $\pi^0\pi^0$ (left) and $\eta\eta$ (right) cross sections integrated over 
the indicated angular ranges and plotted versus $E_{cm}$.
The fits of Figures~\ref{fig:2pi0_csecmfit_bin1_5} and 
\ref{fig:2eta_csecmfit_bin1_7} are re-performed allowing  
$M_{\chi_{c0}}$ and $\Gamma_{\chi_{c0}}$ to be free parameters.
}
\end{figure}
%%%%%%%%%%%%%%%%%%%%%%%%%%%%%%%%%%%%%%%%%%%%%%%%%%%%%%%%%%%%%%%%%%%%%%%%
The results presented so far have been obtained with 
the $\chi_{c0}$ mass ($M_{\chi_{c0}}$) and width ($\Gamma_{\chi_{c0}}$)
constrained to the high-precision values coming from the analysis of the
$J/\psi\,\gamma$ channel \cite{psigamma}.
It is worth checking the capability of the interference technique in 
determining $M_{\chi_{c0}}$ and $\Gamma_{\chi_{c0}}$, in addition to determining the
product of the input and output branching ratios.
The fits shown in Figures~\ref{fig:2pi0_csecmfit_bin1_5} 
and \ref{fig:2eta_csecmfit_bin1_7} are redone in 
Figure~\ref{fig:2pspaper_massgammafree}, with $M_{\chi_{c0}}$ 
and $\Gamma_{\chi_{c0}}$ left as free parameters. 
The results for $M_{\chi_{c0}}$, $\Gamma_{\chi_{c0}}$, 
$B_{in}\times B_{out}$ and $\delta_A$ for both $\pi^0\pi^0$ and $\eta\eta$
channels are given in Table~\ref{tab:e835_results1} in Section~\ref{sec-results}.  

As can be seen, the interferometric technique is able to determine the resonance 
parameters without relying on additional inputs.
The uncertainties on $M_{\chi_{c0}}$ and $\Gamma_{\chi_{c0}}$ from the 
analysis of the $\pi^0\pi^0$ channel alone are 
larger but still comparable to those from the analysis of the  
$J/\psi\,\gamma$ channel.

\section{\label{sec:pi0eta_cross_section}The \bm{$\pi^0\eta$} Cross Section}

\begin{figure*}
\begin{center}
\includegraphics[width=35pc]{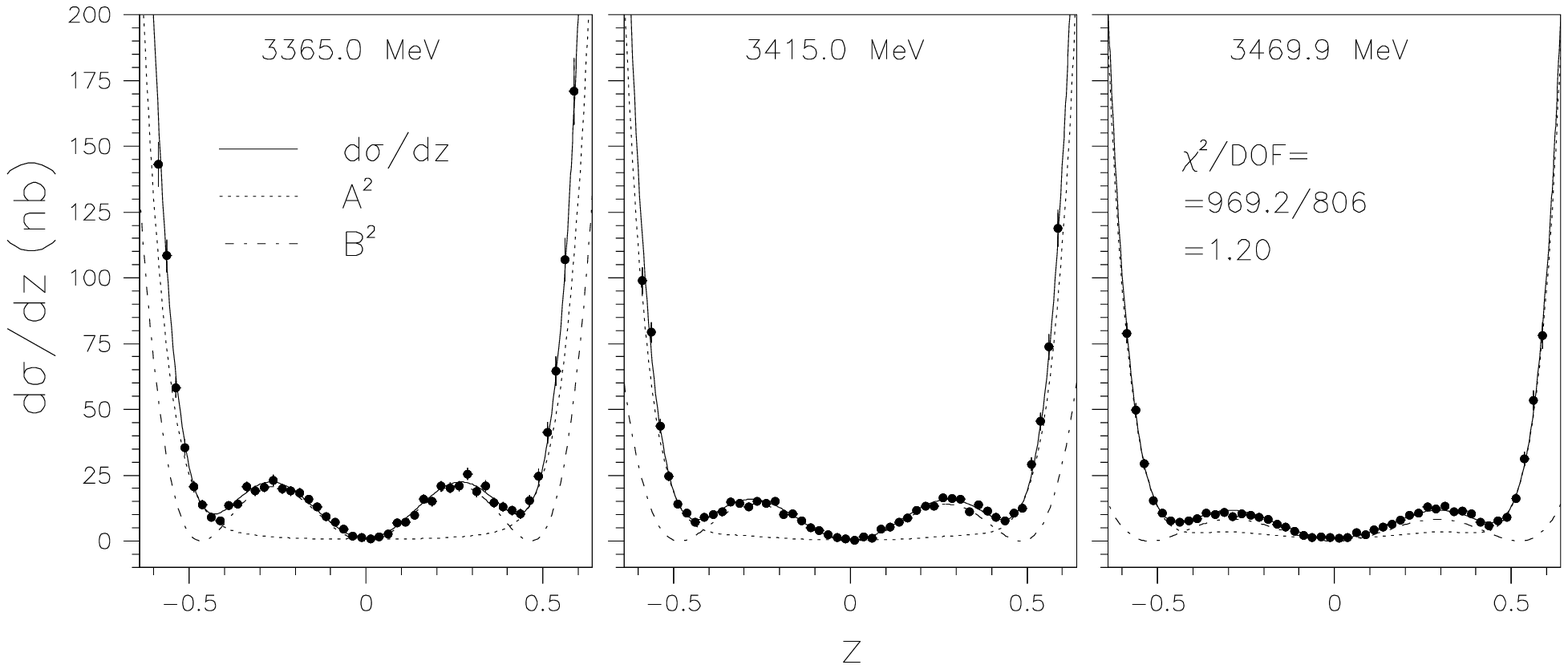}
\end{center}
\caption{         \label{fig:2pspaper_pi0eta_diffcsfit_3stacks} 
The $\bar{p}p\rightarrow\pi^0\eta$ differential cross section versus 
$z\equiv\cos\theta^*_{\pi}$ at different $E_{cm}$. 
A fit using Equations~(\ref{eq:withampres_AandB})-(\ref{eq:noninterferingamp}) 
with $A_R\equiv 0$ is shown along with its components.
The values of the fit parameters are reported in Table~\ref{tab:coefficients_alt}.
}
\end{figure*}

\begin{figure}
\begin{center}
\includegraphics[width=21pc]{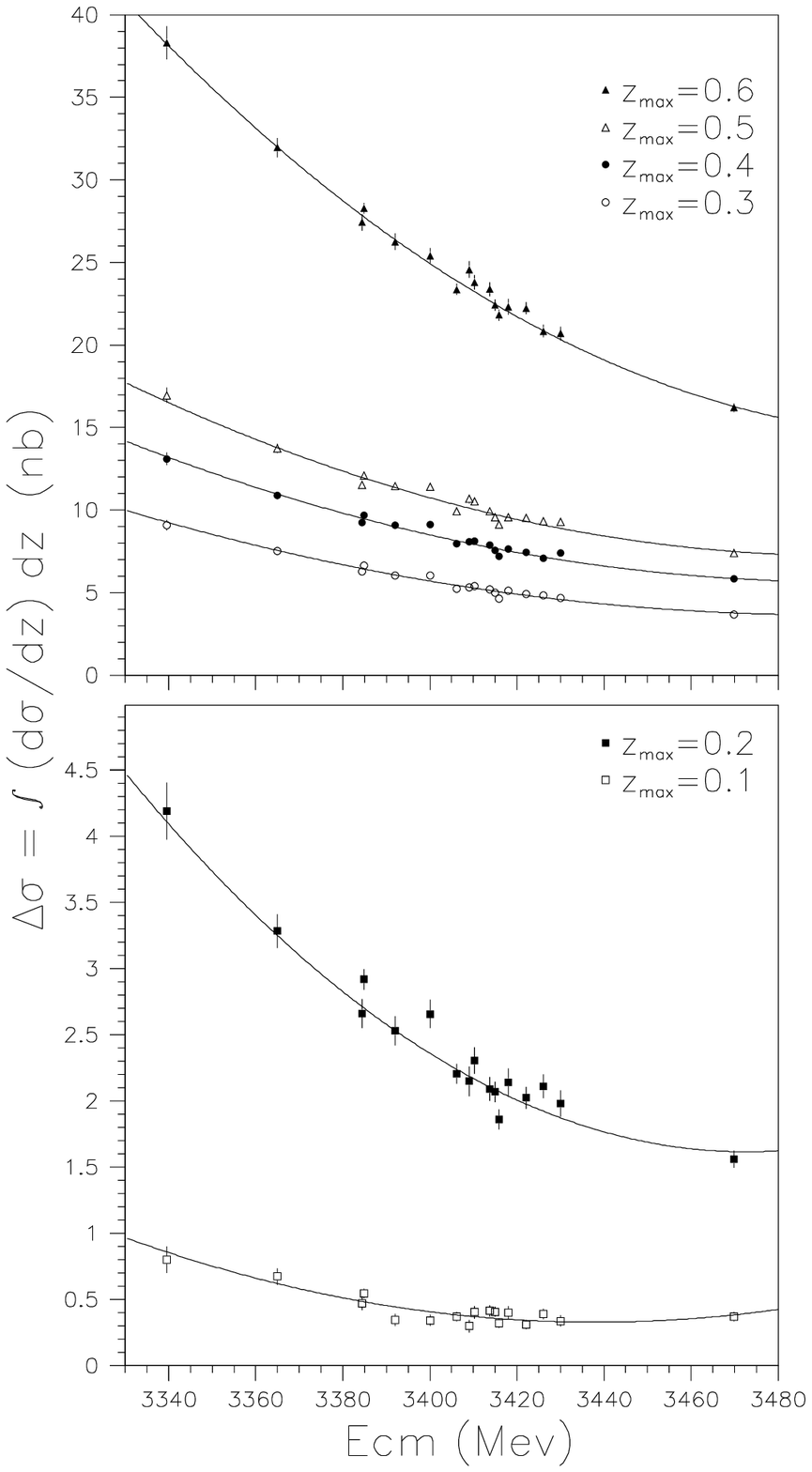}
\end{center}
\caption{         \label{fig:2pspaper_pi0eta_csecm_fit_0_zmax} 
The measured $\bar{p}p\rightarrow\pi^0\eta$ cross section integrated over 
$\vert z\equiv\cos\theta^*_{\pi}\vert <z_{max}$ plotted versus $E_{cm}$.
A fit using Equations~(\ref{eq:withampres_AandB})-(\ref{eq:noninterferingamp}), 
with $A_R=0$, is shown.
}
\end{figure}

The measured $\bar{p}p\rightarrow\pi^0\eta$ differential cross section
is shown in Figure~\ref{fig:2pspaper_pi0eta_diffcsfit_3stacks} for 3 of the 
17 energy points over the angular range 
$-0.6<z\equiv\cos\theta^*_{\pi^0}<0.6$.
The integrated cross section versus $E_{cm}$ is shown in 
Figure~\ref{fig:2pspaper_pi0eta_csecm_fit_0_zmax}; there are 85,751 
background-subtracted $\pi^0\eta$ events.
Also shown is a binned maximum-likelihood fit performed simultaneously on all 
17 energy points to Equation~\ref{eq:withampres_AandB}-\ref{eq:noninterferingamp}. 
The parameter $A_R$ is set to zero, as the decay of any charmonium state 
into $\pi^0\eta$ is suppressed by isospin conservation.

%%%%%%%%%%%%%%%%%%%%%%%%%%%%%%%%%%%%%%%%%%%%%%%%%%%%%%%%%%%%%%%%%%%%%%%%
%%%%%%%%%%%%%%%%%%%%%%%%%%%%%%%%%%%%%%%%%%%%%%%%%%%%%%%%%%%%%%%%%%%%%%%%
%%%%%%%%%%%%%%%%%%%%%%%%%%%%%%%%%%%%%%%%%%%%%%%%%%%%%%%%%%%%%%%%%%%%%%%%
%%%%%%%%%%%%%%%%%%%%%%%%%%%%%%%%%%%%%%%%%%%%%%%%%%%%%%%%%%%%%%%%%%%%%%%%

Since the initial $\bar{p}p$ state is a linear combination of C-eigenstates, 
the final $\pi^0\eta$ angular distribution must be symmetric in the center-of-mass polar angle.
Figure~\ref{fig:2pspaper_bkg_costh} shows that the data, uncorrected for 
acceptance, are obviously not symmetric for $\vert z \vert \gtrsim 0.5$. 
This arises because slow moving $\eta$'s, i.e. those emitted backward in 
the center of mass, have a significantly larger opening angle for the decay photons than
$\pi^{0}$'s of the same center-of-mass angle and thus a greater probability of producing
a photon outside the acceptance, particularly near the acceptance boundaries.
The calculated acceptance is shown 
in \label{pag:symmetrycheck} Figure~\ref{fig:2pspaper_eff_costh}.
Figure~\ref{fig:2pspaper_pi0eta_diffcsfit_3stacks} shows that 
the forward/backward symmetry is recovered in the acceptance-corrected
angular distribution.

The values of the fit parameters are given in Table~\ref{tab:coefficients_alt}. 
The need for the $J=4$ amplitude is more evident than in the $\pi^0\pi^0$ channel
and a fit with $J_{max}=2$ cannot reproduce the multiple minima at 
$z=0$ and $\vert z \vert \simeq 0.45$\,.
The dip in the differential cross section at $z=0$ is very pronounced;
see Figure~\ref{fig:2pspaper_pi0eta_diffcsfit_3stacks}.
As explained before, the helicity-1 component $B^2$ vanishes at $z=0$,
and the helicity-0 component $A^2$ of the cross section is very small 
at $z=0$ due to a local cancellation of the involved partial waves.
The pronounced minimum in the differential cross section is not 
present everywhere within the $E_{cm}$ range 2911~MeV to 3686~MeV~\cite{e760_2body}.

Fits to the cross section are performed with the resonance
amplitude $A_R$ as a free parameter. 
The procedure to obtain the upper limit
\begin{equation}
B(\chi_{c0} \rightarrow \bar{p}p)\times B(\chi_{c0} \rightarrow\pi^0\eta)
<4\times 10^{-8}
\label{eq:pi0eta_BinBout_upperlimit}
\end{equation}
at $90\%$ confidence level is described in detail in \cite{Paolo}. 
This upper limit is one tenth of the values for 
$\pi^{0}\pi^{0}$ and $\eta\eta$  given in Table~\ref{tab:e835_results2}.

\section{\label{sec:etaprime_cross_sections}The \bm{$\pi^0\eta'$} 
and \bm{$\eta\eta'$} Cross Sections}

The small $\eta'\rightarrow\gamma\gamma$ branching ratio 
and a larger background limit the achievable precision of 
the study of channels (4) and (5).
The presence of $\pi^0\eta'$ and $\eta\eta'$ events
can be recognized in Figure~\ref{fig:2pspaper_3ent_scatterplot}.

\begin{figure}
\begin{center}
\includegraphics[width=21pc]{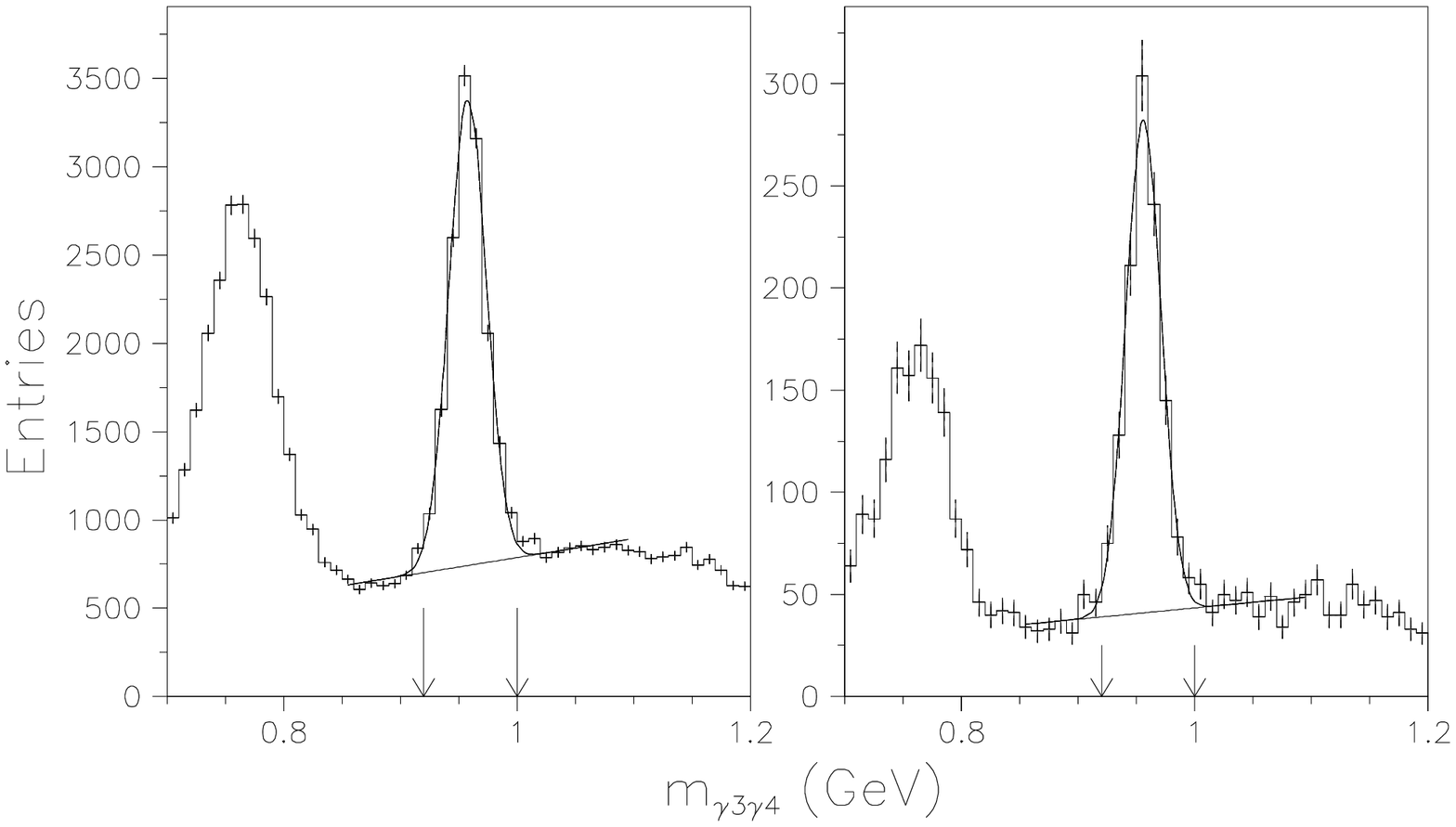}
\end{center}
\caption{         \label{fig:2pspaper_etap_peaks} 
The $\eta'(958)\rightarrow\gamma_3\gamma_4$ peak for the $\pi^0\eta'$ (left) 
and $\eta\eta'$ (right) selections. 
The peak at lower energy is due to 
$\omega(782)\rightarrow\pi^0\gamma,~\pi^0\rightarrow\gamma\gamma$ events
where one of the $\pi^{0}$ decay photons is not observed.
The fit is a gaussian plus a first-degree polynomial.
Arrows indicate the applied cut.
}
\end{figure}

The event selection and variables employed are similar to those described in 
Section~\ref{sec:selection}. 
Events with four CCAL energy deposits greater than 100~MeV are selected
and a 5\% confidence level on a 4C fit 
to $\bar{p}p\rightarrow 4 \gamma$ is required.
The event topology is defined as the combination of the four photons
[named as $\pi^0$~(or~$\eta$)~$\rightarrow\gamma_1\gamma_2$ 
and $\eta'\rightarrow\gamma_3\gamma_4$] with the 
highest confidence level of a 6C fit 
to $\bar{p}p\rightarrow\pi^0\eta'~(\mathrm{or}~\eta\eta') \rightarrow 4 \gamma$.
Coplanarity and colinearity cuts as listed in 
Table~\ref{table-kincuts} are then applied. 
For both analyses, it is additionally required that 
$m_{\gamma_i\gamma_j}>250$~MeV for 
combinations $i,j=1,3;~1,4;~2,3;~\mathrm{and}~2,4$ 
(i.e. the photon pairings not chosen as the event topology) 
to reject $\pi^0$ contamination.
For the $\eta\eta'$ analysis, it is additionally required that 
the sum of the ``wrong'' paired combinations
$m_{\gamma_1\gamma_3}+m_{\gamma_2\gamma_4}$ and
$m_{\gamma_1\gamma_4}+m_{\gamma_2\gamma_3}$ 
are greater than 2.5~GeV. This cut does
not seriously affect the acceptance as seen by MC simulation.
A substantial improvement of the resolution of the $\eta'$ peak 
(from $\sigma\simeq$~40~MeV to 16~MeV) is obtained by using the output values
of the photon energies and positions of a 5C fit to
$\bar{p}p\rightarrow\pi^0~(\mathrm{or}~\eta)~\gamma_3\gamma_4 
\rightarrow\gamma_1\gamma_2~\gamma_3\gamma_4$. 
The resulting distribution of $m_{\gamma_3\gamma_4}$ is shown in 
Figure~\ref{fig:2pspaper_etap_peaks} for $\pi^0\eta'$ and $\eta\eta'$ events.
A cut $\vert m_{\gamma_3\gamma_4}-m_{\eta'}\vert<40$~MeV is applied.
%%%%%%%%%%%%%%%%%%%%%%%%%%%%%%%%%%%%%%%%%%%%%%%%%%%%%%%%%%%%%%%%%%%%%%%%
%%%%%%%%%%%%%%%%%%%%%%%%%%%%%%%%%%%%%%%%%%%%%%%%%%%%%%%%%%%%%%%%%%%%%%%%

\begin{figure}
\begin{center}
\includegraphics[width=21pc]{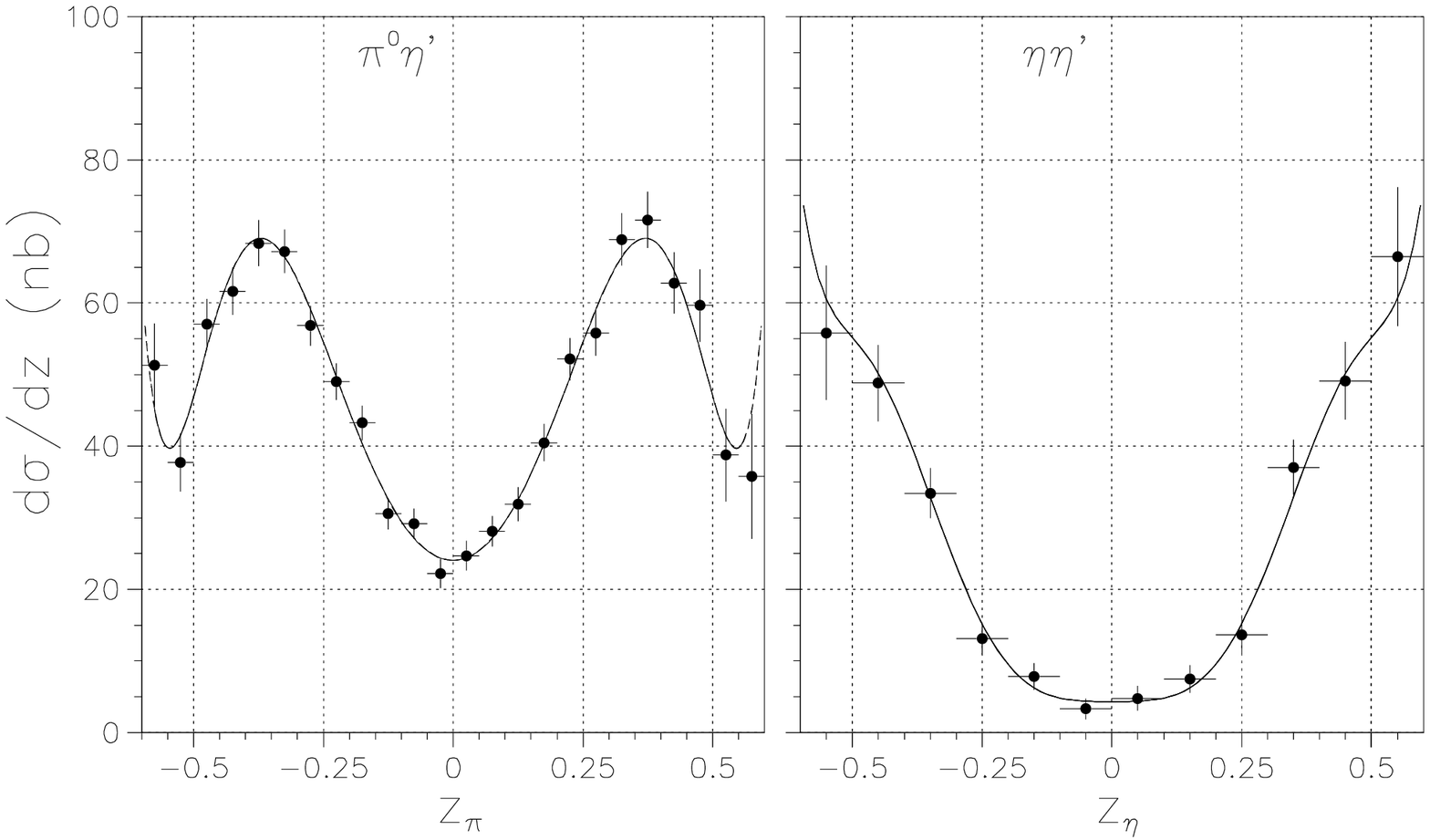}
\end{center}
\caption{         \label{fig:2pspaper_etap_angdis} 
The $\bar{p}p\rightarrow\pi^0\eta'$ and 
$\bar{p}p\rightarrow\eta\eta'$ differential cross sections
at $E_{cm}=3340-3470$~MeV. 
The background has been subtracted.
A fit to a power expansion in even-powers up to $z^8$ (corresponding to
$J_{max}=4$) is shown to verify forward/backward symmetry.
}
\end{figure}

To determine the angular distributions with adequate resolution, 
all 17 energy points are merged together. 
A plot of $m_{\gamma_1\gamma_2}$~versus~$m_{\gamma_3\gamma_4}$ 
(prior to the 5C fit) shows that the
``berm" along the line $m_{\gamma_3\gamma_4}$=$\,m_{\eta'}$ is negligible
compared to the one along $m_{\gamma_1\gamma_2}$=$\,m_{\pi^0}$ 
and along $m_{\gamma_1\gamma_2}$=$\,m_{\eta}$. 
Hence, the background is simply determined by fitting 1-dimensional
projections (after the 5C fit) like those in 
Figure~\ref{fig:2pspaper_etap_peaks}. 
This is done for each bin $\Delta z$, where
$z\equiv\cos\theta^*_{\pi}$ ($z\equiv\cos\theta^*_{\eta}$) 
for the $\pi^0\eta'$ ($\eta\eta'$) analysis.
In the range $\vert z\vert<0.6$ there are 15,097 candidate 
$\pi^0\eta'$ events of which $\sim33\%$ are estimated to be background.
In the range $\vert z\vert<0.6$ there are 1166  
$\eta\eta'$ candidates with an estimated background of $\sim25\%$.
As in the $\pi^0\eta$ case, the acceptances for $\pi^0\eta'$ and 
$\eta\eta'$ are not forward/backward symmetric.
The symmetries are recovered in the background-subtracted and 
efficiency-corrected angular distributions as shown in 
Figure~\ref{fig:2pspaper_etap_angdis}.
%%%%%%%%%%%%%%%%%%%%%%%%%%%%%%%%%%%%%%%%%%%%%%%%%%%%%%%%%%%%%%%%%%%%%%%%
%%%%%%%%%%%%%%%%%%%%%%%%%%%%%%%%%%%%%%%%%%%%%%%%%%%%%%%%%%%%%%%%%%%%%%%%

\begin{figure}
\begin{center}
\includegraphics[width=21pc]{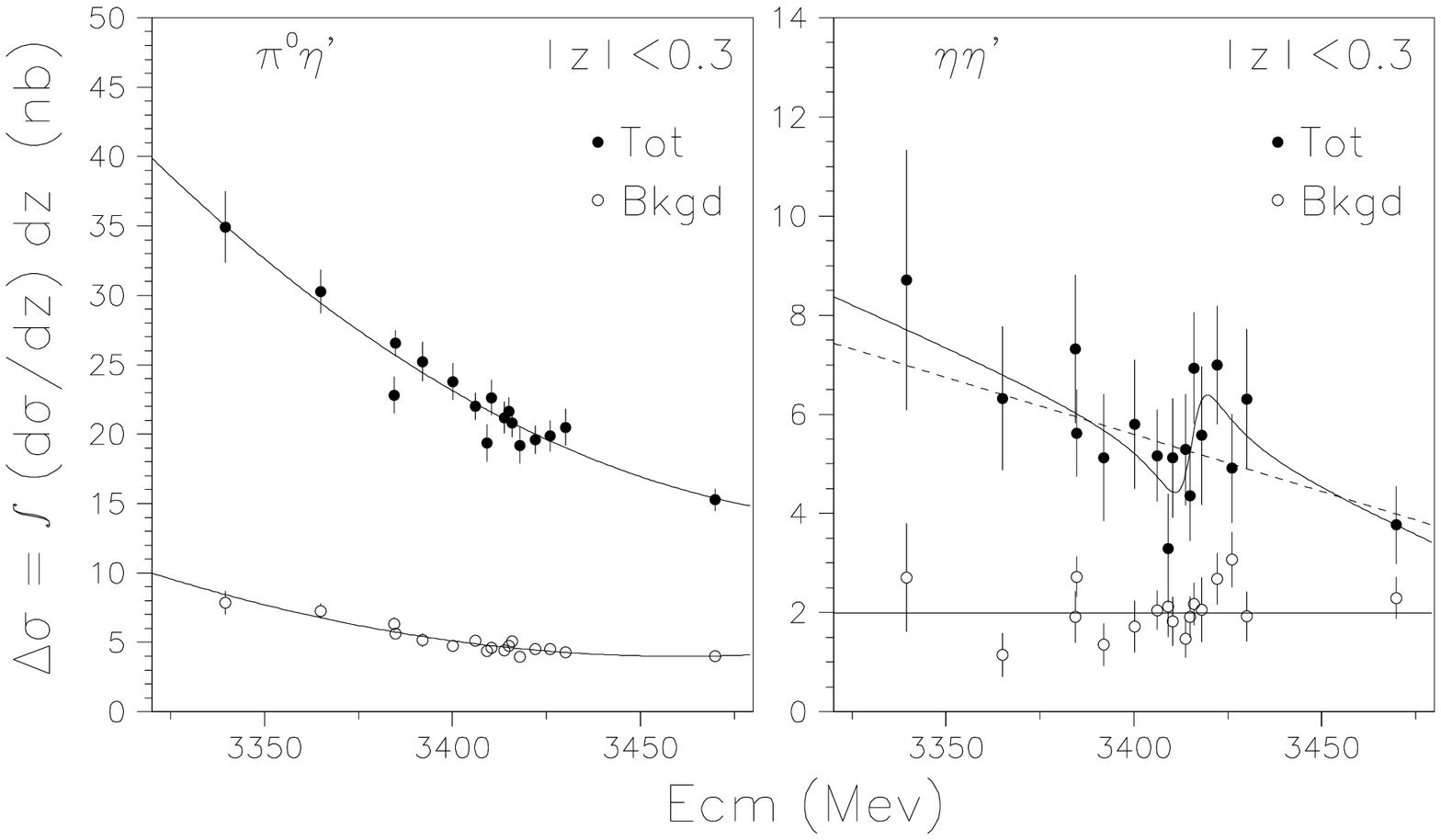}
\end{center}
\caption{         \label{fig:2pspaper_etap_scans} 
The $\bar{p}p\rightarrow\pi^0\eta'$ and $\bar{p}p\rightarrow\eta\eta'$ 
cross sections integrated over $\vert z \vert <0.3$ 
($z\equiv\cos\theta^*_{\pi}$ and $z\equiv\cos\theta^*_{\eta}$
for $\pi^0\eta'$ and $\eta\eta'$, respectively) 
plotted versus $E_{cm}$.
In these two plots, the amount of cross section due to background (Bkgd) 
has not been subtracted from the total (Tot), and the fits shown 
are performed simultaneously on Tot and Bkgd cross sections (see text).}
\end{figure}

Figure~\ref{fig:2pspaper_etap_scans} shows the
$\bar{p}p\rightarrow\pi^0\eta'$ and 
$\bar{p}p\rightarrow\eta\eta'$ cross sections integrated for
$\vert z \vert<0.3$ versus $E_{cm}$ 
(the background has not been subtracted in the plot). 
In practice, the background has
been determined at each energy point with fits like those in 
Figure~\ref{fig:2pspaper_etap_peaks}.
The data are not sufficient to perform partial-wave 
analyses to distinguish the helicity-0 and helicity-1 components of the
continuum and the fits in figure~\ref{fig:2pspaper_etap_scans} are performed
simply on the integrated cross section.
Equation~\ref{eq:interfterm} is used.
Given the small values of $\vert z \vert$ in these fits, 
the helicity-1 component $B^2$ is fixed to zero for both channels. 
The background is parameterized as a polynomial
and fit simultaneously with the total cross section. 
%%%%%%%%%%%%%%%%%%%%%%%%%%%%%%%%%%%%%%%%%%%%%%%%%%%%%%%%%%%%%%%%%%%%%%%%
%%%%%%%%%%%%%%%%%%%%%%%%%%%%%%%%%%%%%%%%%%%%%%%%%%%%%%%%%%%%%%%%%%%%%%%%

Charmonium decay into $\pi^0\eta'$ ($\eta\eta'$) violates (satisfies) 
isospin conservation.
For the $\pi^0\eta'$ channel, the quality of the fit remains unaltered by 
allowing the resonant amplitude $A$ free or fixing it to zero.
The $\chi^2$ is 23.9 in both cases and the degrees of freedom 
are 27 or 29, respectively. 
The $90\%$ confidence level upper limit is 
\begin{equation}
B(\chi_{c0} \rightarrow \bar{p}p)\times B(\chi_{c0} \rightarrow\pi^0\eta')
< 2.5 \times 10^{-7}.
\label{eq:pi0etap_BinBout_upperlimit}
\end{equation}
For the $\eta\eta'$ channel, the fit to the total cross section changes 
from the dashed to the solid line in Figure~\ref{fig:2pspaper_etap_scans} 
when a resonant amplitude $A$ is allowed. 
The $\chi^2/DOF$ goes from 30.4/31 to 25.6/29. 
If the feature observed in the data is due to interference between 
the $\chi_{c0}$ and the continuum, the solid line fit would imply 
\begin{equation}
B(\chi_{c0} \rightarrow \bar{p}p)\times B(\chi_{c0} \rightarrow\eta\eta')
= (2.1^{+2.3}_{-1.5}) \times 10^{-6}.
\label{eq:etaetap_BinBout}
\end{equation}
The error quoted is statistical and dominates the systematic uncertainty.

\section{\label{sec-results}Results}

\subsection{The \bm{$\chi_{c0}$} State}

The E835 measurements of the $\chi_{c0}$ resonance are summarized
in Tables~\ref{tab:e835_results1} and~\ref{tab:e835_results2}.
%The E835 measurements of the $\chi_{c0}$ resonance in the channels
%    $~\bar{p}p\rightarrow J/\psi\,\gamma$, $J/\psi\rightarrow e^+e^-~$;
%    $~\bar{p}p\rightarrow \pi^0\pi^0 \rightarrow 4 \gamma~$;  
%and $~\bar{p}p\rightarrow \eta\eta \rightarrow 4 \gamma~$
%are summarized in Table~\ref{tab:e835_results}. An upper limit
%from the isospin-violating channel 
%$~\bar{p}p\rightarrow \chi_{c0}\rightarrow\pi^0\eta \rightarrow 4 \gamma~$
%is provided in Equation~(\ref{eq:pi0eta_BinBout_upperlimit}).   
%The smaller samples of events 
%    $~\bar{p}p\rightarrow \pi^0\eta' \rightarrow 4 \gamma~$ and  
%    $~\bar{p}p\rightarrow \eta\eta' \rightarrow 4 \gamma~$
%give the results in Equations~(\ref{eq:pi0etap_BinBout_upperlimit})
%and~(\ref{eq:etaetap_BinBout}).
%The results from the channels $J/\psi\,\gamma$ and $\pi^0\pi^0$
%have been already published in 
%\cite{psigamma} and \cite{2pi0}, respectively.
The results from the channels $J/\psi\,\gamma$ and $\pi^0\pi^0$
have been already published in 
\cite{psigamma} and \cite{2pi0}, respectively.

\begin{table*}
\caption{ \label{tab:e835_results1}
E835 Results for the $\chi_{c0}$ with $M_{\chi_{c0}}$, $\Gamma_{\chi_{c0}}$, 
and $B_{in}\times B_{out}$ as free parameters.}
\begin{tabular}{|c|c|c|c|}\hline
$~~~~~~~~~~~~~B_{in}\equiv$		&\multicolumn{3}{c|}{Common channel $B(\chi_{c0}\rightarrow \bar{p}p)$}	     \\ 
$~~~~~~~~~~~~B_{out}\equiv$		&$B(\chi_{c0}\rightarrow J/\psi\,\gamma)$		&$B(\chi_{c0}\rightarrow \pi^0\pi^0)$	&$B(\chi_{c0}\rightarrow \eta\eta)$	    \\ \hline
%\hline
$M_{\chi_{c0}}$ (MeV/c$^2$)		&$3415.4\pm0.4\pm0.2$      
&$3414.7^{+0.7}_{-0.6}\pm0.2$    	&$3412.2^{+2.1}_{-1.8}\pm0.2$                   \\
$\Gamma_{\chi_{c0}}$ (MeV)		&$9.8\pm1.0\pm0.1$         
&$8.6^{+1.7}_{-1.3}\pm0.1$       	&$10.3^{+3.0}_{-3.1}\pm0.1$                     \\
$B_{in} \times B_{out}$ ($10^{-7}$)	&$27.2\pm1.9\pm1.3$  
&$5.42^{+0.91}_{-0.96}\pm0.22$    	&$4.1^{+1.2}_{-1.1}\,^{+0.5}_{-0.3}$          \\ 
$\delta_A$ (degree)	      		&$-$        		           
&$47\pm10\pm6$               			&$173^{+17}_{-19}\pm6$                         	\\ 
\hline
\end{tabular}
\end{table*}
%%%%%%%%%%%%%%%%%%%%%%%%%%%%%%%%%%%%%%%%%%%%%%%%%%%%%%%%%%%

The header section of Table~\ref{tab:e835_results1} indicates the common formation
channel ($\bar{p}p$) and the different $\chi_{c0}$ decay channels:
$J/\psi\,\gamma$, $\pi^0\pi^0$, and $\eta\eta$.
The bottom part of the table reproduces the results for 
$M_{\chi_{c0}}$, $\Gamma_{\chi_{c0}}$, and $B_{in}\times B_{out}$
as determined by each of the three analyses independently.
For $\pi^0\pi^0$ and $\eta\eta$, the phase $\delta_A$ at small $z$ 
between the helicity-0 nonresonant and resonant amplitudes is also given.
There is good agreement for the values of $M_{\chi_{c0}}$ 
and $\Gamma_{\chi_{c0}}$ from the $J/\psi\,\gamma$ and $\pi^0\pi^0$ channels.
The limited sample of $\eta\eta$ events gives lower-precision measurements;
the width is in agreement with the other two channels, while the mass
is slightly underestimated (correlated with a probable slight 
overestimate of the phase $\delta_A$).
The background from misidentified events from different processes
is negligible for all three cases.
There is an absence of nonresonant production 
in the $J/\psi\ \gamma$ channel.
The large nonresonant continuum in the $\pi^0\pi^0~$ and 
$\eta\eta$ channels, although useful
as the provider of the amplification for the interference pattern,  
requires additional parameters to be fit. 
In particular, the phase $\delta_A$ and the size of the interfering continuum
increase the coupling among the fit parameters.
As a result, the statistical uncertainties are larger than in the 
$J/\psi\,\gamma$ channel.

Table~\ref{tab:e835_results2} presents
the results for $B_{in} \times B_{out}$ and the phase $\delta_A$
as determined for the neutral pseudoscalar channels
by constraining $M_{\chi_{c0}}$ and $\Gamma_{\chi_{c0}}$
to the measurements from the  $J/\psi\,\gamma$ channel.
The $J/\psi\,\gamma$ data sample was collected
at the same time as the four photon events.
The energy and luminosity determinations are the same for all these channels.
By using the  values of $M_{\chi_{c0}}$ and $\Gamma_{\chi_{c0}}$
from the $J/\psi\,\gamma$ analysis, 
the values of $B_{in} \times B_{out}$ and $\delta_A$ in the
$\pi^0\pi^0$~channel are determined with higher accuracy and precision.
The values in  Table~\ref{tab:e835_results2}
are our final results for $B_{in} \times B_{out}$ and $\delta_A$.
%%%%%%%%%%%%%%%%%%%%%%%%%%%%%%%%%%%%%%%%%%%%%%%%%%%%%%%%%%%
\begin{table*}
\caption{ \label{tab:e835_results2}
E835 Results for the $\chi_{c0}$ with $M_{\chi_{c0}}$ and $\Gamma_{\chi_{c0}}$ 
constrained to the values of the $J/\psi\,\gamma$ results in Table~\ref{tab:e835_results1}.}
\begin{tabular}{|c|c|c|c|c|c|}\hline
$~~~~~~~~~~~~~B_{in}\equiv$		&\multicolumn{5}{c|}{Common channel $B(\chi_{c0}\rightarrow \bar{p}p)$}	     \\ 
$~~~~~~~~~~~~B_{out}\equiv$		&$B(\chi_{c0}\rightarrow \pi^0\pi^0)$	&$B(\chi_{c0}\rightarrow \eta\eta)$	     &$B(\chi_{c0}\rightarrow \eta\eta')$&$B(\chi_{c0}\rightarrow \pi^0\eta)$	&$B(\chi_{c0}\rightarrow \pi^0\eta')$\\ 
\hline
%\multicolumn{2}{|c|}{Final result for $B_{in} \times B_{out}$ ($10^{-7}$)}
$B_{in} \times B_{out}$ ($10^{-7}$)
 &$5.09\pm0.81\pm0.25$			&$4.0\pm1.2^{+0.5}_{-0.3}$   &$21^{+23}_{-15}$  & $< 0.4~(90\% CL)$ & $< 2.5~(90\% CL)$ \\
%\multicolumn{2}{|c|}{Final result for phase $\delta_A$ (degree)}
$\delta_A$ (degree)
 &$39\pm5\pm6$	      			&$144\pm8\pm6$					     &$189\pm20$ &$-$ &$-$\\
\hline
\end{tabular}
\end{table*}

The value of the phase $\delta_A$ between the 
helicity-0 nonresonant and resonant amplitudes
is determined in a restricted angular region at small $z$ 
($z<0.125$ for $\pi^0\pi^0$, $z<0.35$ for $\eta\eta$
and $z<0.3$ for $\eta\eta'$).
No appreciable energy dependence is seen within the considered regions of $z$.
The phase $\delta_A$ is responsible for the shape of the interference
pattern seen in the cross section and is reasonably well measured.
Its specific value is determined by the local combination of several partial 
waves (see Equation~\ref{eq:interferingamp}), which largely cancel each other.

In the ratio between $B_{in} \times B_{out}$ of two analyzed channels, the
common $B_{in}$ and some systematics cancel out. 
We obtain
%%%%%%%%%%%%%%%%%%%%%%%%%%%%%%%%%%%%%%%%%%%%%%%%%%%%%%%%%%%%%%%%%%%%%%%%%
\begin{equation}
\frac{B(\chi_{c0}\rightarrow \eta\eta)}
{B(\chi_{c0}\rightarrow \pi^0\pi^0)}
=0.79\pm0.27^{+0.10}_{-0.06}~.
\label{eq:B2pi0_B2eta}
\end{equation}
%%%%%%%%%%%%%%%%%%%%%%%%%%%%%%%%%%%%%%%%%%%%%%%%%%%%%%%%%%%%%%%%%%%%%%%%%
The BES experiment has reported the measurement of
$B(\chi_{c0}\rightarrow \pi^0\pi^0)=(2.79\pm0.32\pm0.57)\times 10^{-3}$,
which agrees with the 1985 measurement from Crystal Ball~\cite{PDG}.
However, a consistent and more precise determination of this quantity 
may be computed by using isospin symmetry: 
$B(\chi_{c0}\rightarrow\pi^0\pi^0)=\frac{1}{2}B(\chi_{c0}
\rightarrow\pi^+\pi^-)=(2.5\pm0.35)\times 10^{-3}$,
where $B(\chi_{c0}\rightarrow\pi^+\pi^-)=(5.0\pm0.7)\times 10^{-3}$ 
is taken from \cite{PDG}.
This and Equation~\ref{eq:B2pi0_B2eta} provide
%%%%%%%%%%%%%%%%%%%%%%%%%%%%%%%%%%%%%%%%%%%%%%%%%%%%%%%%%%%%%%%%%%%%%%%%%
\begin{equation}
B(\chi_{c0}\rightarrow \eta\eta)=
(1.98\pm0.68_{stat}\,^{+0.25}_{-0.15\,sys}\pm0.28_{PDG})\times 10^{-3}~,
\label{eq:B_chi0_2eta}
\end{equation}
%%%%%%%%%%%%%%%%%%%%%%%%%%%%%%%%%%%%%%%%%%%%%%%%%%%%%%%%%%%%%%%%%%%%%%%%%
where the subscript PDG labels the uncertainty derived from errors listed 
in~\cite{PDG}.
This is in agreement with a measurement
$B(\chi_{c0}\rightarrow \eta\eta)=(2.02\pm0.84\pm0.59)\times 10^{-3}$ 
reported by BES \cite{BES_chi0_2pi0}, 
again in agreement with a 1985 measurement from Crystal Ball \cite{PDG}.

\subsection{Non-resonant \bm{$\bar{p}p$} Annihilation  
into Two Pseudoscalar Mesons}

In addition to the charmonium results, this work provides fits to
the cross sections for three pseudoscalar-pseudoscalar meson states
in antiproton-proton annihilations in the energy 
range $3340$~MeV~$-3470$~MeV (see Table~\ref{tab:coefficients_alt}).
The differential cross section provides insights on the dynamics 
of these processes at the examined energies.
The differential cross sections as a function of the production angle of the 
meson pair are shown in 
Figures~\ref{fig:2pspaper_2pi0_diffcsfit_3stacks},
\ref{fig:2pspaper_pi0eta_diffcsfit_3stacks} and 
\ref{fig:2pspaper_2eta_diffcsfit_3groups}
for $\pi^0\pi^0$, $\pi^0\eta$ and $\eta\eta$, respectively, and
in Figure~\ref{fig:2pspaper_etap_angdis} for $\pi^0\eta'$ and $\eta\eta'$.
Tables with the numerical values for 
$\pi^0\pi^0$, $\pi^0\eta$ and $\eta\eta$ can be found in \cite{Paolo}.

The partial-wave expansion fits described in 
Sections~\ref{sec:2pi0_cross_section} ($\pi^0\pi^0$), 
\ref{sec:pi0eta_cross_section} ($\pi^0\eta$) 
and \ref{sec:2eta_cross_section} ($\eta\eta$) 
indicate that the size of the angular-momentum 
contribution decreases rapidly with $J$
and only $J=0,\,2$~and~$4$ are necessary to describe the angular distributions.
The statistical resolution for $\pi^0\eta'$ and $\eta\eta'$ is
limited. However, their angular distributions are well fitted by 
an even-power expansion up to $z^8$, hence are consistent with $J_{max}=4$. 

Applying the semi-classical relation $~\hbar \, L_{\bar{p}p}\simeq b \, p_i~$
(where $L_{p\bar{p}}$ is the orbital angular momentum of the 
$\bar{p}p$ system, $b$ is the impact parameter, and $p_i$ is the initial 
state center-of-mass momentum), it can be inferred
that a larger number of partial waves would participate.
This is probably true for other reactions and is known to be true for
elastic $\bar{p}p$ scattering.
However, the pseudoscalar-pseudoscalar meson states selected here are
unlikely to be produced in peripheral impacts, since they require the 
annihilation of one, two or three valence $q\bar{q}$ pairs of 
the $\bar{p}p$ initial state. 
If only one $q\bar{q}$ pair annihilates, the remaining valence quarks of the 
proton ($qq$) and antiproton ($\bar{q}\bar{q}$) must separate and rearrange
themselves into $~q\bar{q}+q\bar{q}$\,.
This is unlikely to happen at large impact parameters, where the partons 
tend to retain their original large longitudinal components of momentum.
An indirect observation of a sharply decreasing trend in the
relative contribution of increasing $L_{\bar{p}p}$ is provided by the
exclusive process $\bar{p}p\rightarrow c\bar{c}$, which necessarily requires
total valence-quark annihilation. 
The ratios $\Gamma_{\bar{p}p}/\Gamma_{\mathrm{gluons}}$ for the 
charmonium states
that couple to $L_{\bar{p}p}=0$ ($\eta_c$, $J/\psi$ and $\psi'$)
are 5-10 times larger than for those that couple to
$L_{\bar{p}p}=1$ (the $\chi_{cJ}$'s) \cite{PDG}.
Table~\ref{tab:quantumnumbers} indicates that only odd values 
of $L_{\bar{p}p}$ can produce the $~J^{PC}=\mathrm{even}^{++}~$ 
of a pseudoscalar meson pair,
and each $L_{\bar{p}p}$ feeds both $~J=L_{\bar{p}p}\pm1~$.
It is then reasonable to expect larger contributions from the partial waves
with $J=0$~and~2 as compared to  $J=4$ as 
observed in Table~\ref{tab:coefficients_alt}.
It may also be noticed that the estimated differences of phase 
between amplitudes with the same helicity are relatively small.

\section{Discussion and Conclusions}

E835 has studied the formation of the $\chi_{c0}$ state of charmonium 
in antiproton-proton annihilation and its subsequent decay into 
pseudoscalar-pseudoscalar mesons.
In the $\pi^0\pi^0$ and $\eta\eta$ channels, an 
interference-enhanced pattern is evident in a cross section
dominated by the nonresonant production of 
pairs of pseudoscalar mesons.

The choice of performing this study on the $\chi_{c0}$~resonance is 
a consequence of the $~J^{PC}=0^{++}~$ quantum numbers of the $\chi_{c0}$,
which allows it to decay into a pseudoscalar meson pair.
The primary goal of E835 was the determination of the resonance parameters
through the study of the process 
$~\bar{p}p\rightarrow J/\psi\,\gamma$, $J/\psi\rightarrow e^+e^-~$
to complete the program of studying the $\chi_{cJ}$ 
triplet initiated by E760.
Thanks to the antiproton source developments mentioned in the 
introduction, a large integratged luminosity was collected in the $\chi_{c0}$~region 
in the year 2000 run of E835.
The presence of a large nonresonant continuum cross section for 
the final $\pi^0\pi^0$ and $\eta\eta$ states, 
as compared to the resonant production through a charmonium
intermediary state, was known at the outset.
However, the awareness that the interference mechanism would produce 
an enhanced interference pattern in the cross section, 
hopefully large enough to be detected, was a strong motivation to pursue 
the analysis.
Of course, the physics behind the interference mechanism was well known 
prior to this work. 
What is innovative is the exploitation of such a  
mechanism to detect and measure a resonant signal that would, 
if the interference could be turned off, be almost two orders of magnitude 
smaller than the nonresonant cross section.
Several other measurements of interference patterns exist in particle physics 
and many in nuclear physics, but the usual experience is with 
a larger resonance amplitude and a smaller interfering continuum.
This specific analysis requires the correct separation of the two different
continuum components, one interfering and the other not interfering with the 
resonance. 
The effectiveness of this analysis has been demonstrated.

A reason for pursuing this study
was the search for alternative means of discovering and measuring 
charmonium states and possible hadromolecular states.
Now that confidence has been gained that measurements of 
resonances can be accomplished in hadronic decay channels where
the nonresonant production of the final state dominates,
new strategies can be considered.
For example, other than the above mentioned search for hadromolecules,
the study of the poorly known charmonium singlet states 
could be performed by investigating hadronic final states, relying on the
enhancement provided by the interference.

\begin{acknowledgments}
% put your acknowledgments here.
The authors thank the staff of their institutions and in particular 
the Antiproton Source Department of the Fermilab Accelerator Division.
This research was supported by the US Department of Energy and 
the Italian Istituto Nazionale di Fisica Nucleare.
\end{acknowledgments}

\end{document}